\documentclass[aps,twocolumn,english,superscriptaddress,showpacs,longbibliography]{revtex4-1}
\usepackage{amsmath}
\usepackage{amssymb,scalerel}
\usepackage{graphicx}
\usepackage[colorlinks,bookmarks=false,citecolor=red,linkcolor=blue,urlcolor=blue]{hyperref}
\usepackage{lipsum, color}
\usepackage{wasysym}
\def\Tr{\mathop{\mathrm{Tr}}} 
\def\hc{\mathrm{H.c.}}
\def\Hhat{\hat{H}}

\def\i{{\boldsymbol i}}
\def\j{{\boldsymbol j}}
\def\k{{\boldsymbol k}}

\def\Q{{\boldsymbol Q}}

\newcommand{\ve}[1]{\boldsymbol{#1}}
\begin{document}
\title{Zooming in on heavy fermions in  Kondo lattice models}
\author{Bimla Danu}
\affiliation{Institut f\"ur Theoretische Physik und Astrophysik and W\"urzburg-Dresden Cluster of Excellence ct.qmat, Universit\"at W\"urzburg, 97074 W\"urzburg, Germany}
\author{Zihong Liu}
\affiliation{Institut f\"ur Theoretische Physik und Astrophysik and W\"urzburg-Dresden Cluster of Excellence ct.qmat, Universit\"at W\"urzburg, 97074 W\"urzburg, Germany}
\author{Fakher F. Assaad}
\affiliation{Institut f\"ur Theoretische Physik und Astrophysik and W\"urzburg-Dresden Cluster of Excellence ct.qmat, Universit\"at W\"urzburg, 97074 W\"urzburg, Germany}
\author{Marcin Raczkowski}
\affiliation{Institut f\"ur Theoretische Physik und Astrophysik, Universit\"at W\"urzburg, 97074 W\"urzburg, Germany}

\date{\today}
\begin{abstract}
Resolving the  heavy fermion band  in the  conduction electron momentum resolved spectral  function of the Kondo lattice model  is challenging since, in the weak coupling limit,  its spectral  weight  is  exponentially  small.   In this article we  consider a   composite  fermion operator,  consisting of a 
conduction electron dressed by spin fluctuations  that shares the same quantum  numbers as the 
electron operator.  
Using approximation free auxiliary field  quantum Monte Carlo simulations we   show that for the SU(2) spin-symmetric  model on the  square lattice at half filling,  the  composite fermion acts as a magnifying glass for the  heavy fermion band.  In comparison to the  conduction electron residue that scales as $e^{-W/J_k}$  with $W$ the bandwidth and $J_k$   the  Kondo coupling,  the  residue of the   composite fermion  tracks $J_k$.    This  result  holds   down to $J_k/W = 0.05$,   and  confirms the point  of view  that   magnetic  ordering,   present below $J_k/W = 0.18$,  does  not  destroy  the  heavy  quasiparticle. 
We furthermore  investigate the  spectral  function of the composite fermion in the ground state and at finite temperatures,   
for SU($N$) generalizations of the  Kondo lattice model, as well as for   ferromagnetic Kondo couplings, and compare  our  results to analytical calculations in  the limit of  high  temperatures,  large-$N$,  large-$S$, and large  $J_k$. Based on these calculations,   we conjecture that  the composite  fermion operator   provides  a unique tool  to  study the  destruction of the heavy fermion 
quasiparticle in Kondo  breakdown  transitions.  The relation of our  results to  scanning tunneling spectroscopy   and photoemission experiments  is  discussed. 
\end{abstract}

\maketitle
\section{Introduction} 
A   quasiparticle   excitation  is defined by its quantum numbers,  spin-$1/2$,  unit charge,   and  crystal momentum for the electron,   and infinite lifetime.   
This leads to  the   generic form of the  single particle  retarded  Green's function:
\begin{equation}
    G(\ve{k}, \omega)  =    \frac{Z_{\ve{k}}}{   \omega  + i 0^{+} - E(\ve{k}) }   +  G^{inc}(\ve{k},\omega),
\end{equation}
where $Z_{\ve{k}}$  is the quasiparticle  residue, $E(\ve{k}) $  the dispersion relation and   $G^{inc} $ the incoherent  background. 
Generically,  one  expects    that any  operator,  $ \hat{\psi}^{\dagger}_{\ve{k},\sigma} $,  carrying  the same quantum numbers as the quasiparticle,   to   reveal the same  dispersion  relation.  However the quasiparticle   residue,  $ Z_{\ve{k}} \equiv   |\langle   \Psi_{0}^{n+1} (\ve{k}) |   \hat{\psi}^{\dagger}_{\ve{k},\sigma} | \Psi_{0}^{n} \rangle |^2$,    corresponding to the overlap between the  ground state   in the $n$-particle sector with an additional   quasiparticle of momentum $\ve{k}$   and the ground state in the  $n+1$ particle sector and  and momentum  $\ve{k}$,   will depend on  the  specific form of the operator $\hat{\psi}^{\dagger}_{\ve{k},\sigma}$.     For  example,   for an  antiferromagnetic insulator,  the quasiparticle should be  understood in terms of a fermion dressed  with spin fluctuations,   a spin polaron \cite{PhysRevB.44.317,PhysRevB.62.15480},  or  alternatively  by a  bound state of  a spinon and  holon \cite{BERAN1996707, Grusdt18}.   
 If  $Z_{\ve{k}} $ is small,  then measuring  the Green's function of the electron    may not be an optimal  strategy.   A  workaround is to  optimize the  specific form of $ \hat{\psi}^{\dagger}_{\ve{k},\sigma} $  so as to  maximize  $Z_{\ve{k}}$~\cite{PhysRevLett.76.279}.     Such an approach is appealing since,  provided that single particle  excitations   exist,  it allows  one  to \textit{ zoom in}  on them and  understand  the nature of the 
 \textit{dressing} 
   of the bare electron.         
The    failure to   find an operator  $  \hat{\psi}^{\dagger}_{\ve{k},\sigma}  $   with $Z_{\ve{k}}  > 0$  is even more interesting since it  signals the  breakdown of  the  quasiparticle  picture.    In one-dimensions, this is  generic  \cite{Giamarchi}.  In higher dimensions, notions  such as orthogonal  metals  with  fractionalized  fermions   can  be put  forward~\cite{Senthil03,Hohenadler18,Hohenadler19}. 

In this article, we  will concentrate on the heavy fermion state as realized for example in CeCu$_6$~\cite{Lohneysen_rev}.   These materials can be understood  in terms of a lattice  of magnetic impurities,  stemming from  the localized Ce-4$f$  electrons,   embedded in a metallic host.   In the local moment regime where  charge  fluctuations of the Ce-4$f$ electron  can safely be omitted,  the   adequate model   to  describe these materials is the Kondo lattice model (KLM),  with  Kondo coupling $J_k$  between the localized  spins and  the  spin degree  of freedom of the  conduction electrons. CeCu$_6$   has an effective mass that   exceeds  by many orders the magnitude of the bare electron mass.    Numerical  simulations \cite{Assaad04a} as well as large-$N$    calculations of the KLM \cite{Burdin00} show that  the enhancement of  the effective mass stems  from the frequency dependence  of the  self-energy  of the  bare electron  
retarded Green's  function.   In particular, the quasiparticle  residue  in the small  $J_k/W$ limit  tracks the  Kondo   scale   $ Z_{\ve{k}}  \simeq e^{-W/J_k} $. Here $W$ corresponds to the  bandwidth.

The  question we will ask here  is if we can   define a fermion operator $ \hat{\ve{\psi}}^{\dagger}_{\ve{k}} $  that enhances the spectral  weight  of the heavy  fermion band.   Let us first assume that  the KLM  can be derived from a periodic Anderson model (PAM)  describing   the same conduction electron band hybridizing  with a narrow $f$-band.  In  the local moment  regime, a  canonical  Schrieffer-Wolff~\cite{Schrieffer66} transformation provides a mapping between  both  models.    In the paramagnetic phase~\cite{Capponi00},  the conduction electron spectral function will  exhibit heavy bands  but  with very low spectral weight.    Since the heavy  band  has $f$-character,  one should   actually  consider the $f$-single particle spectral function  to resolve it.     The mapping between the Kondo lattice and periodic Anderson models,   provides a  simple scheme to derive the  fermion operator that one should compute in the realm of the KLM to resolve the heavy band. It merely  corresponds to the  Schrieffer-Wolff canonical transformation of the $f$-fermion operator  in the PAM~\cite{Raczkowski18}.   For a  conduction $\hat{\ve{c}}^{\dagger}_{\i} $  and impurity spin   $\hat{\ve{S}}_\i$  in the unit cell $\i$, it reads:
\begin{equation}
	\hat{\ve{\psi}}^{\dagger}_{\i} =  \hat{\ve{c}}^{\dagger}_{\i} \ve{\sigma}  \cdot  \hat{\ve{S}}^{}_\i, 
\label{composite_fermion}
\end{equation}
and corresponds to the form put forward in Refs.~\cite{Costi00,Maltseva09}.  
This   fermion operator is  relevant for the understanding of scanning tunneling microscopy (STM) spectra of magnetic adatoms on metallic 
surfaces~\cite{Toskovic2016,Moro2019,Raczkowski18,Danu19,Morr19} and Kondo lattice materials~\cite{PRL.104.187202,PRL.105.246401,Aynajian2010,Yazdani12}. In particular,  within the single impurity  Kondo model, it reveals the Kondo  resonance~\cite{Costi00,Raczkowski18}.   Anderson and Appelbaum used  $\hat{\ve{\psi}}^{\dagger}_{\i} $ to explain the zero-bias tunneling anomalies in $s$-$d$ exchange models~\cite{AndersonPW1966,PhysRevAppelbaum1967,PRLAppelbaumJ1966}.  The aim of this article is to take the step  from  the impurity to the lattice and  compute, with approximation free quantum Monte Carlo methods,  the  momentum resolved  spectral function of the 
composite fermion.  

The    richness of  phenomena that  can  be captured by  considering the   fermion operator $\hat{\ve{\psi}}^{\dagger}_{\i}$ is remarkable   and can be investigated   by considering   several limiting cases.    First of all,   $\hat{\ve{\psi}}^{\dagger}_{\i}$ is a composite object of   a spin and fermion degrees of freedom. Hence if  one neglects  interactions between  these two entities,  the spectral  function   $A_{\psi}(\ve{k}, \omega) $  will  show a broad  continuum  of excitations corresponding  to the convolution of the conduction electron spectral function and spin susceptibility of the   impurity spins  (see  Sec.~\ref{vrtx_cntrbts}).     Poles in the spectral function of  the composite fermion  operator correspond to  bound states  of  spins and conduction electrons. In fact, in the zero temperature  and  large-$N$ limits~\cite{Raczkowski20}  one will show  that  $A_{\psi}(\ve{k}, \omega) $  exhibits quasiparticle  poles akin to the  hybridized band picture of heavy  fermions.  The  weight of  these  poles is  proportional    to  the    square of the  hybridization mean-field order  parameter  $V$ (see Sec.~\ref{Large-N}).  Hence, within this approximation,  the  Kondo  breakdown   transition,   characterized by $V=0$,    is revealed by the   vanishing of the   heavy  fermion pole in  the composite  fermion spectral function.  Momentum integrated calculations of this quantity  for a spin chain on a  semi-metallic surface,  support this point of view~\cite{Danu20}.   
 In the strong coupling limit, $A_{\psi}(\ve{k}, \omega) $    can be computed  and  equally  shows a pole structure  (see  Sec.~\ref{Strong_cpl}). 
Furthermore, one  fundamental  question in the realm of  heavy fermion systems is the fate of Kondo screening in the  magnetically   ordered  phase  triggered  by the  Ruderman-Kittel-Kasuya-Yosida (RKKY) interaction~\cite{Ruderman1954,Kasuya56,Yosida1957}.  At  the mean-field   level   the question   boils  down to a finite or vanishing value of the hybridization mean-field order  parameter  $V$,   that is   again   revealed  by  the  
weight of the  quasiparticle pole in $A_{\psi}(\ve{k}, \omega) $. 
Finally, if   Kondo screening  is not present in   the magnetically  ordered  phase,  one can  adopt a large-$S$ approximation  (see  Sec.~\ref{largeSthry}).   In leading order in $S$,  the  spectral  function  $A_{\psi}(\ve{k}, \omega) $  will correspond  to the conduction electron spectral  function   shifted by the ordering wave vector  $\ve{Q}$.

The organization of the article is as follows.  In  Sec.~\ref{modelham} we introduce the Kondo lattice Hamiltonian.   
Section~\ref{psif_op}  is devoted to  a  detailed  discussion of the fermion $\hat{\ve{\psi}}^{\dagger}_{\i}$    operator.   We  will first  discuss its symmetry properties  and then make predictions 
concerning its spectral properties based on the high temperature, large-$N$~\cite{Burdin00,hazra2021luttinger}, strong 
coupling~\cite{PhysRevB.55.R6109,PhysRevB.58.7599}, and large-$S$ limits. 
In Sec.~\ref{QMC_method}  we summarize the details of the auxiliary field   quantum Monte Carlo (QMC) simulations.   
In Sec.~\ref{QMC_results}  we present our QMC results for the SU(2)  and SU($N$)   antiferromagnetic Kondo lattice  models as  well as  
for the SU(2)  ferromagnetic KLM.   In  Sec.~\ref{Conclusions}   we   conclude and provide outlooks.

\section{Model Hamiltonian} \label{modelham}
\begin{figure}[htbp]
\centering
\includegraphics[width=0.49\textwidth]{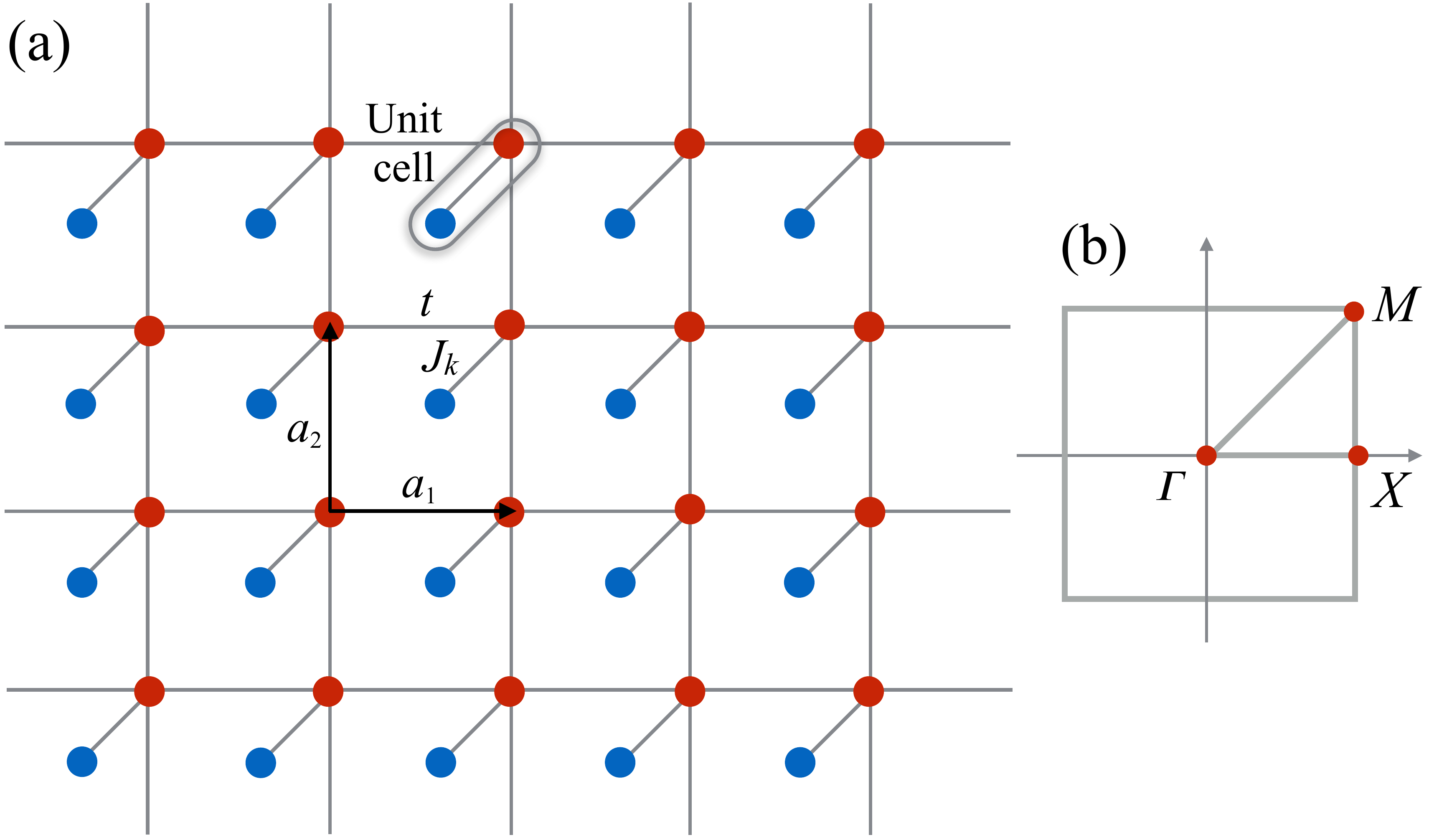}
\caption{(a) Sketch of the square KLM with the nearest neighbor hopping $t$ and  Kondo exchange coupling $J_k$: 
the conduction and localized orbitals are indicated by red and blue dots. (b)  
First Brillouin zone and its high symmetry points: $\Gamma=(0,0)$, $X=(\pi,0)$, and $M=(\pi,\pi)$.}
\label{Lattice_SKLM}
\end{figure}

\begin{figure}[htbp]
\includegraphics[width=0.47\textwidth]{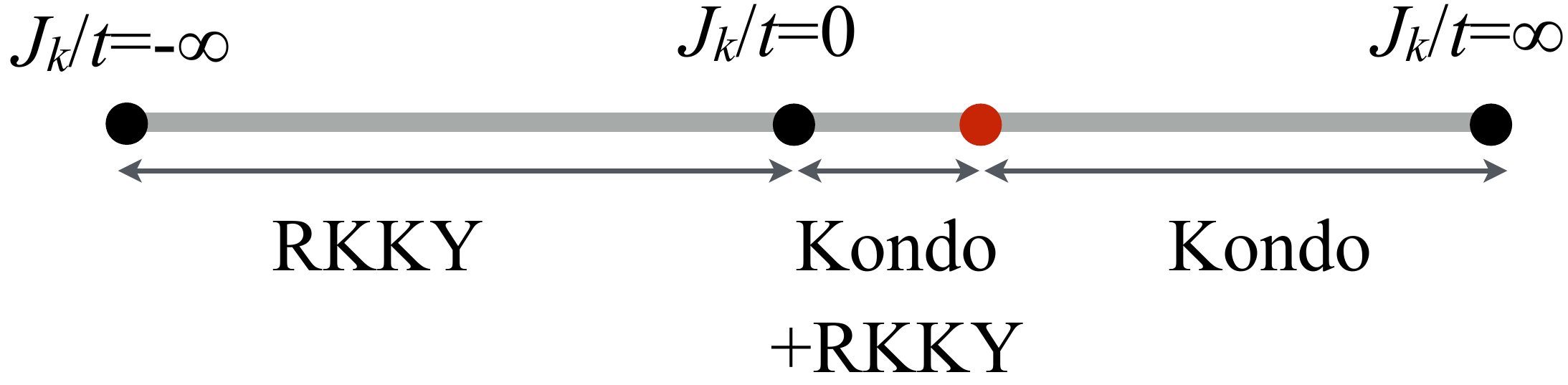}
\caption{Conjectured ground state phase diagram for a square KLM at half filling.  
For $J_k/t>0$, a quantum critical point (red dot) separates the Kondo-screened and antiferromagnetically ordered phases. 
In the latter, the composite fermion spectral function is consistent with the coexistence of Kondo screening and the RKKY interaction; 
for $J_k/t<0$, the RKKY interaction is the only relevant energy scale.}
\label{Phases_SKLM}
\end{figure}

We start with the Kondo lattice Hamiltonian 
\begin{eqnarray}
 \Hhat_{KLM} &= &\sum_{ \k,\sigma} \epsilon_{\k}{\hat c}^{\dagger}_{\k,\sigma} { \hat c}^{}_{\k,\sigma} + 
        J_k   \sum_{\i} \hat{\ve S}^c_{\i} \cdot \hat{\ve S}^{}_{\i}.
 \label{model_ham}
\end{eqnarray}
Here, the operator ${\hat c}^{\dagger}_{\k,\sigma}$ creates an electron with a wave vector $\k$ and $z$ component of spin $1/2(-1/2)$,  
$\epsilon_\k$  describes the band dispersion energy, and  $J_k$ is the Kondo exchange coupling between conduction electron spins 
$\hat{\ve S}^c_{\i} =\frac{1}{2} \sum_{\sigma \sigma^\prime} 
{\hat c}^\dagger_{\i, \sigma}   \ve{\sigma}_{\sigma, \sigma^\prime} {\hat c}_{\i, \sigma^\prime}$ 
and localized magnetic moments 
$\hat{\ve S}^{}_{\i}  $ 
with $\ve{\sigma}$  being  the Pauli matrices. 
Specifically, we have  considered  a square KLM with the hopping amplitude $t$ restricted to nearest neighbors, see Fig.~\ref{Lattice_SKLM}. 
Our aim is  elucidate the response of  the composite fermion operator in different parts of the  phase diagram illustrated  in Fig.~\ref{Phases_SKLM}. 

 \section{Composite fermion operator}\label{psif_op}
 To  introduce  the composite fermion operator, it is convenient to take a step back and assume that the  KLM   can be derived from a periodic  Anderson  model (PAM):
 \begin{eqnarray}
 \Hhat_{PAM} &= &\sum_{ \k,\sigma} \epsilon_{\k}{\hat c}^{\dagger}_{\k,\sigma} { \hat c}^{}_{\k,\sigma}
 + V\sum_{\i, \sigma } \big(   \hat c^{\dagger}_{\i,\sigma}    \hat f^{\phantom\dagger}_{\i,\sigma}    +  \text{H.c} \big)     \nonumber \\
& + &  \frac{U}{2}  \sum_{\i}  \big(  \hat{n}_{\i}^{f} -1  \big)^2. 
 \label{PAM}
\end{eqnarray}
The Hamiltonian $\hat{H}_{KLM}$ in Eq.~(\ref{model_ham}) is obtained in the limit of  strong Hubbard interaction $U$ on the $f$ orbitals 
by carrying out a   canonical Schrieffer-Wolff~\cite{Schrieffer66}  transformation  of the PAM, $\hat{H}_{PAM}$.  
Hence, $e^{\hat{S}} $$\hat{H}_{PAM} $$ e^{-\hat{S}}   = \hat{H}_{KLM}$  with  $\hat{S}^\dagger = - \hat{S}$.    
Then,   the composite fermion operator is given by: 
\begin{equation}
	e^{\hat{S}} \hat{f}^{\dagger}_{\i,\sigma'} e^{-\hat{S}}   \simeq  
     \frac{2V}{U}  \left( {\hat c}^{\dagger}_{\i,-\sigma'}  \hat{S}^{\sigma'}_{\i} +  \sigma'   {\hat c}^{\dagger}_{\i,\sigma'} \hat{S}^{z}_\i   \right)  \equiv  \frac{2V}{U} 	\hat{\psi}^{\dagger}_{\i,\sigma'}.   
\end{equation}
In the above, it is understood that $\sigma'$ takes the value $1$ ($-1$)  for up  (down) spin degrees of freedom, that  
$  \hat{S}^{\sigma'}_{\i} =  {\hat f}^{\dagger}_{\i,\sigma'} {\hat f}^{}_{\i,-\sigma'}  $  and that 
$ \hat{S}^{z}_{\i} = \frac{1}{2} \sum_{\sigma'}  \sigma' {\hat f}^{\dagger}_{\i,\sigma'} {\hat f}^{}_{\i,\sigma'} $.    
This form matches that derived in Ref.~\cite{Costi00} and a calculation  of the former equation can be found in Ref.~\cite{Raczkowski18}.  
An equivalent, but more transparent formulation is given in  Ref.~\cite{Maltseva09} and reads: 
\begin{equation}
		\hat{\psi}^{\dagger}_{\i,\sigma}= \sum_{\sigma'} \hat{c}^{\dagger}_{\i,\sigma'} \ve{\sigma}^{}_{\sigma',\sigma} \cdot  \hat{\ve{S}}^{}_\i
\end{equation}
where $\ve{\sigma}$  denotes  the vector of Pauli spin matrices.  

\subsection{Symmetry properties of the composite fermion}
As the composite fermion operator  stems from a canonical  transformation of the ${f}$-fermion operator, it must share   identical symmetry properties.     However, since the canonical transformation was  carried  out   within perturbation theory,  the  statement is not  exact  and   a  calculation will  show that the  anti-commutation rules for the composite  fermion operator   read: 
\begin{eqnarray}
	  & &    \big\{ {\hat{\psi}}^{\dagger}_{\ve{i},\sigma},  {\hat{\psi}}^{\phantom\dagger}_{\ve{j},\sigma'}  \big\} =\delta_{\ve{i},  \ve{j} }  \big(  i \hat{\ve{S}}_{\ve{i}}   \cdot  \big[ \big(  \hat{\ve{c}}_{\ve{i}}^{\dagger} \ve{\sigma} \big)_{\sigma}   \times    
	            \big( \ve{\sigma}  \hat{\ve{c}}_{\ve{i}}^{\phantom\dagger} \big)_{\sigma'}  \big] \nonumber  \\
	   & &-  \ve{\sigma}_{\sigma,\sigma'}  \cdot \hat{\ve{S}}_{\ve{i}}  +  
	            \delta_{\sigma,\sigma'} S(S+1)   \big) 
\end{eqnarray}
and 
\begin{eqnarray}
	            	 \big\{ {\hat \psi}^{\dagger}_{\ve{i},\sigma}, {\hat \psi}^{\dagger}_{\ve{j},\sigma'}  \big\}  =   \delta_{\ve{i},  \ve{j} }  
	  \big(  i \hat{\ve{S}}_{\ve{i}}   \cdot  \big[ \big(\hat{\ve{c}}_{\ve{i}}^{\dagger} \ve{\sigma} \big)_{\sigma}   \times    
	            \big( \ve{\sigma}  \hat{\ve{c}}_{\ve{i}}^{\dagger} \big)_{\sigma'}  \big]  \big). 
\end{eqnarray}
The above holds  for the   spin $S=1/2$ case. 
As a consequence the sum rule  for a   composite fermion   spectral  function   defined  as 
\begin{equation}\label{Afkomegaeq}
	   A_{\psi} ( \k,\omega)    =  - \frac{1}{\pi} \text{Im}    G^{\text ret}_{\psi}(\k,\omega) 
\end{equation}
with 
\begin{equation}\label{Gfkomegaeq}
 G^{\text ret}_{\psi}(\k,\omega)=-i \int^\infty_0 dt e^{i \omega t} \sum_{\sigma} \big \langle \big\{ \hat{{\psi}}_{\k,\sigma}(t),\hat{ {\psi}}^\dagger_{\k,\sigma}(0) \big\}  \big\rangle 
\end{equation}
reads:
\begin{eqnarray}
\label{EQ.sum_rule}
\int d \omega  A_{\psi} ( \k,\omega) &  & =    \sum_{\sigma}  \big{\langle} \big\{   \hat{{\psi}}^{\dagger}_{\ve{k},\sigma},  \hat{{\psi}}^{\phantom\dagger}_{\ve{k},\sigma}  \big\}     \big{\rangle}
         \\  
    && = - \frac{2}{N_u} \sum_{\ve{i}} \big\langle  \hat{\ve{S}}_{\ve{i}}   \cdot \hat{\ve{c}}^{\dagger}_{\ve{i}}   \ve{\sigma} \hat{\ve{c}}^{\phantom\dagger}_{\ve{i}}    \big\rangle   + 2S(S+1) \nonumber . 
\end{eqnarray}
Since   $  -3/2 < \langle  \hat{\ve{S}}_{\ve{i}}   \cdot \hat{\ve{c}}^{\dagger}_{\ve{i}}   \ve{\sigma} \hat{\ve{c}}^{}_{\ve{i}}    \rangle    <  1/2 $  the    sum rule   is,  as expected,  
positive and   is  maximal for the antiferromagnetic     alignment of impurity and conduction electron spins. 
Hence  the sum  rule  lies in the interval  $\left[0.5, 4.5\right] $  and is hence very comparable to that of the conduction electron  that takes a  value of two.  

We  now   show that the composite  fermion transforms  as an SU(2)  spinor  under global spin  rotations. 
The  generator  for  global spin  rotations   corresponds to the total  spin:
\begin{equation}
	\hat{\ve{S}}_{\text{tot}}    =  \sum_{\ve{i}}   \big(  \hat{\ve{S}}_{\ve{i}}    +  \frac{1}{2}    \hat{\ve{c}}^{\dagger}_{\ve{i}}  \ve{\sigma} 
	  \hat{\ve{c}}^{\phantom\dagger}_{\ve{i}}   \big) 
\end{equation} 
such that  for 
\begin{equation}
	  \hat{U}(\ve{e},\theta) =  e^{- i \theta  \ve{e}\cdot \hat{\ve{S}}_{\text{tot}}}
\end{equation}
with  $\ve{e}$  a  unit vector  in $\mathbb{R}^{3}$  and $\theta $  a real  angle, 
\begin{equation}
	 \hat{U}^{-1} (\ve{e},\theta)  \hat{\ve{c}}^{\dagger}_{\ve{i}}  \hat{U}(\ve{e},\theta)    =     \hat{\ve{c}}^{\dagger}_{\ve{i}} e^{ i \frac{\theta}{2} \ve{e}\cdot \ve{\sigma} } 
\end{equation}
and 
\begin{equation}
	\hat{U}^{-1} (\ve{e},\theta) \hat{\ve{S}}_{\ve{i}} \hat{U}(\ve{e},\theta)    =     R(\ve{e},\theta)  \hat{\ve{S}}_{\ve{i}}. 
\end{equation}
In the above,  $R(\ve{e},\theta) $  is an  SO(3)  rotation around axis $\ve{e}$   with angle $\theta$.    Since, 
$ e^{ i \frac{\theta}{2} \ve{e}\cdot \ve{\sigma} }      \ve{\sigma}    e^{ -i \frac{\theta}{2} \ve{e}\cdot \ve{\sigma} }     =     R(\ve{e},\theta)  \ve{\sigma} $,  one will show that  $ \hat{\ve{\psi}}^{\dagger}_{\ve{i}}$ transforms as an SU(2) spinor: 
\begin{equation}
	 \hat{U}^{-1} (\ve{e},\theta)  \ve{\hat{\psi}}^{\dagger}_{\ve{i}}\hat{U}(\ve{e},\theta)    =     \ve{\hat{\psi}}^{\dagger}_{\ve{i}} e^{ i \frac{\theta}{2} \ve{e}\cdot \ve{\sigma} } . 
\end{equation}
	    
We will now discuss the behavior of the composite fermion spectral function  upon  neglecting vertex corrections in the large-$N$ and large-$S$  limits. 

\subsection{Omission of vertex  contributions}\label{vrtx_cntrbts}

Omitting vertex corrections,   the composite  fermion  spectral  function   is given by a convolution of the   spin   susceptibility and  the single particle  spectral  function.    In particular,  along the   imaginary time,   the \textit{ bubble } contribution to the  correlation function reads: 
\begin{equation}
\label{Eq.bubble}
          	 \sum_{\sigma} \big\langle  {\hat \psi}^{\phantom\dagger}_{\ve{i},\sigma} {\hat \psi}^{\dagger}_{\ve{j},\sigma}(\tau)     \big\rangle     =      
          \sum_{\sigma}   \big\langle   \hat{\ve{S}}_{\ve{i}}    \hat{\ve{S}}_{\ve{j}}(\tau)     \big\rangle  
          \big \langle   \hat{c}^{\phantom\dagger}_{\ve{i},\sigma}   \hat{c}^{\dagger}_{\ve{j},\sigma}(\tau) \big\rangle.
\end{equation}
Transforming to  real time   and momentum space gives
\begin{eqnarray}
   	A_{\psi} ( \k,\omega)    =  & &\frac{1}{N_u} \sum_{\ve{p}} \int d \Omega A_{c} ( \ve{p},\omega)  \chi''(\ve{k} -  \ve{p},  \omega -  \Omega)    \times \nonumber  \\ 
	& &  \left[   n_B(\omega - \Omega) -  n_F(\Omega) \right].
\end{eqnarray}
In the above,     $\chi''( \ve{k} -  \ve{p},  \omega -  \Omega )  $   corresponds to  the imaginary part  of   the    impurity spin  susceptibility  and  $A_{c} ( \ve{p},\omega)  $ is the spectral function of the conduction electrons.      $n_B(\omega - \Omega) $,   $n_F(\omega) $  correspond respectively to the  Bose-Einstein  and    Fermi-Dirac  distributions  at the  considered   temperature.      Generically, the above  convolution  should   yield   a    broad     composite  fermion spectral  function.    Consider  for example  a  temperature scale    where the    spins  are  disordered   such that the  dynamical  spin  structure factor 
\begin{equation}
	   S(\ve{q}, \omega)   =    \chi''(\ve{q},  \omega ) n_B(\omega)
\end{equation}
can  be approximated by   $  S(\ve{q}, \omega)  \propto  \delta(\omega) $,  such that  $ \chi''(\ve{q},  \omega ) =0   \, \, \forall  \, \,  \omega \neq 0 $.       In this case, 
\begin{equation}\label{Apsieq_kinpndt}
	A_{\psi} ( \k,\omega)      \propto \frac{1}{N_u} \sum_{\ve{p}}  A_{c} ( \ve{p},\omega), 
\end{equation}
  is  $\ve{k}$-independent and  corresponds  to the density of states of  the conduction electrons.    We will see that   our  high temperature  QMC data reproduce 
this form.

\subsection{SU($N$) generalization and the large-$N$ limit}\label{Large-N}

Here  we generalize the SU(2) Kondo lattice model   to SU($N$) in the  totally antisymmetric self-conjugate representation.  
Let $T^{a}$ be the $N^2 -1$ generators of SU($N$) that satisfy the normalization condition: 
\begin{equation}
	\Tr  \left[ T^{a} T^{b} \right]   = \frac{1}{2}\delta_{a,b}.
\label{Normalization_condition.eq}
\end{equation}
For the SU(2) case, $T^{a}$  corresponds to the $\hat{T}=\frac{1}{2} \ve{\sigma}$ with $\ve{\sigma}$   a vector of the three Pauli spin matrices.     
The SU($N$)  generalization of  the KLM then reads: 
\begin{eqnarray}
 \hat{H}^{N}_{KLM} &= &\sum_{ \k,\sigma=1}^{N} \epsilon_{\k}{\hat c}^{\dagger}_{\k,\sigma} { \hat c}_{\k,\sigma}
 +\frac{2J_k}{N}\sum_{\i,a}   \hat{T}_{\i} ^{a,c}\hat{T}_{\i}^{a,f}.
 \label{KLM_SUN}
\end{eqnarray} 
where
\begin{equation}
	 \hat{T}^{a,c}_{\i}   =   \sum_{\sigma,\sigma'=1}^{N} {\hat{c}}^{\dagger}_{\i,\sigma}{T}^{a}_{\sigma,\sigma'}  {\hat{c}}^{\phantom\dagger}_{\i,\sigma'}, \; \; 
	  \hat{T}^{a,f}_{\i}   = \sum_{\sigma,\sigma'=1}^{N} {\hat{f}}^{\dagger}_{\i,\sigma} T^{a}_{\sigma,\sigma'} {\hat{f}}^{\phantom\dagger}_{\i,\sigma'}.  
\end{equation}
The fermionic representation of the SU($N$) generators as well as the constraint
\begin{equation}
   \sum_{\sigma=1}^{N}  \hat{f}^{\dagger}_{\i,\sigma}   \hat{f}^{\phantom\dagger}_{\i,\sigma}  \equiv  \hat{n}^{f}_\i = \frac{N}{2}
\label{constraint}
\end{equation}
define the  self-adjoint  totally antisymmetric representation. 
To  formulate the large-$N$ approximation,  we use the relation:
\begin{equation}
	\sum_{a} T^{a}_{\alpha,\beta} T^{a}_{\alpha',\beta'} = \frac{1}{2} \Big(  \delta_{\alpha,\beta'}  \delta_{\alpha',\beta} - \frac{1}{N} \delta_{\alpha,\beta} \delta_{\alpha', \beta'} \Big), 
\label{SUN_eq}
\end{equation}
to  show that  in the constrained Hilbert space,
\begin{align}
	 \frac{2 J_k}{N} \sum_{ a=1  }^{N^2 -1} & \hat{T}^{a,c}_{\i}  \hat{T}^{a,f}_{\i}   =    - \frac{J_k}{2N} \sum_{\i}  \big( 
                \hat{D}^{\dagger}_{\i} \hat{D}^{\phantom\dagger}_{\i}   +    \hat{D}^{\phantom\dagger}_{\i} \hat{D}^{\dagger}_{\i}    \big)  + \frac{J_k}{4}             
 \end{align}
with
\begin{equation*}
	   \hat{D}^{\dagger}_{\i}   = \sum_{\sigma=1}^{N} \hat{f}^{\phantom\dagger}_{\i,\sigma}  \hat{c}^{\dagger}_{\i,\sigma}  .
\end{equation*}
In the large-$N$ limit,   $\hat{D}^{\dagger}_{\i}$  is of order $N$  and fluctuations around  the mean-field are of order one,  such that the square of the fluctuations  can be  neglected.  The constraint,  $ \sum_{\sigma=1}^{N} \hat{f}^{\dagger}_{\i,\sigma}   \hat{f}^{\phantom\dagger}_{\i,\sigma}  $  is equally of order $N$ and since fluctuations around the mean are  again of order one, it can be imposed on average.  

The composite fermion operator is readily generalized to SU($N$) as: 
\begin{equation}
		{\hat{\psi}}^{\dagger}_{\i,\sigma} = \frac{4}{N} \sum_{\sigma' = 1}^{N} \sum_{a = 1}^{N^2-1}\hat{c}^{\dagger}_{\i,\sigma'} T^{a}_{\sigma',\sigma} \hat{T}^{a,f}_{\i} 
\end{equation}
where the pre-factor has to be chosen such that the above equation matches Eq.~(\ref{composite_fermion})  at $N=2$.   
Using Eq.~(\ref{SUN_eq})  and the constraint, we  obtain
\begin{equation}
	\hat{\psi}^{\dagger}_{\i,\sigma} =   \frac{2}{N}  \big[ \hat{f}^{\dagger}_{\i,\sigma} 
	\big(  \hat{D}^{\dagger}_{\i} - \hat{f}^{\phantom\dagger}_{\i,\sigma} \hat{c}^{\dagger}_{\i,\sigma}  \big) 
        -\hat{c}^{\dagger}_{\i,\sigma} \big(\hat{f}^{\dagger}_{\i,\sigma}\hat{f}^{\phantom\dagger}_{\i,\sigma}  -1/2  \big) \big].
\end{equation}
In the large-$N$  limit,   $  \hat{D}^{\dagger}_{\i} $    scales as $N$  and  fluctuations  of order  one can be neglected such   that
\begin{equation}\label{Veqpsi}
	{\hat{\psi}}^{\dagger}_{\i,\sigma}    \propto  \hat{f}^{\dagger}_{\i,\sigma} 
	 \frac{2}{N} \big< \hat{D}^{\dagger}_{\i}  \big>.
\end{equation} 
In the heavy  fermion  state  characterized by $ \big< \hat{D}^{\dagger}_{\i}  \big>  \neq 0 $  we expect    the composite fermion 
 operator  to  resolve  the  heavy  fermion band since it is of  dominant  $f$-character.

\begin{figure}[t]
\centering
\includegraphics[width=0.46\textwidth]{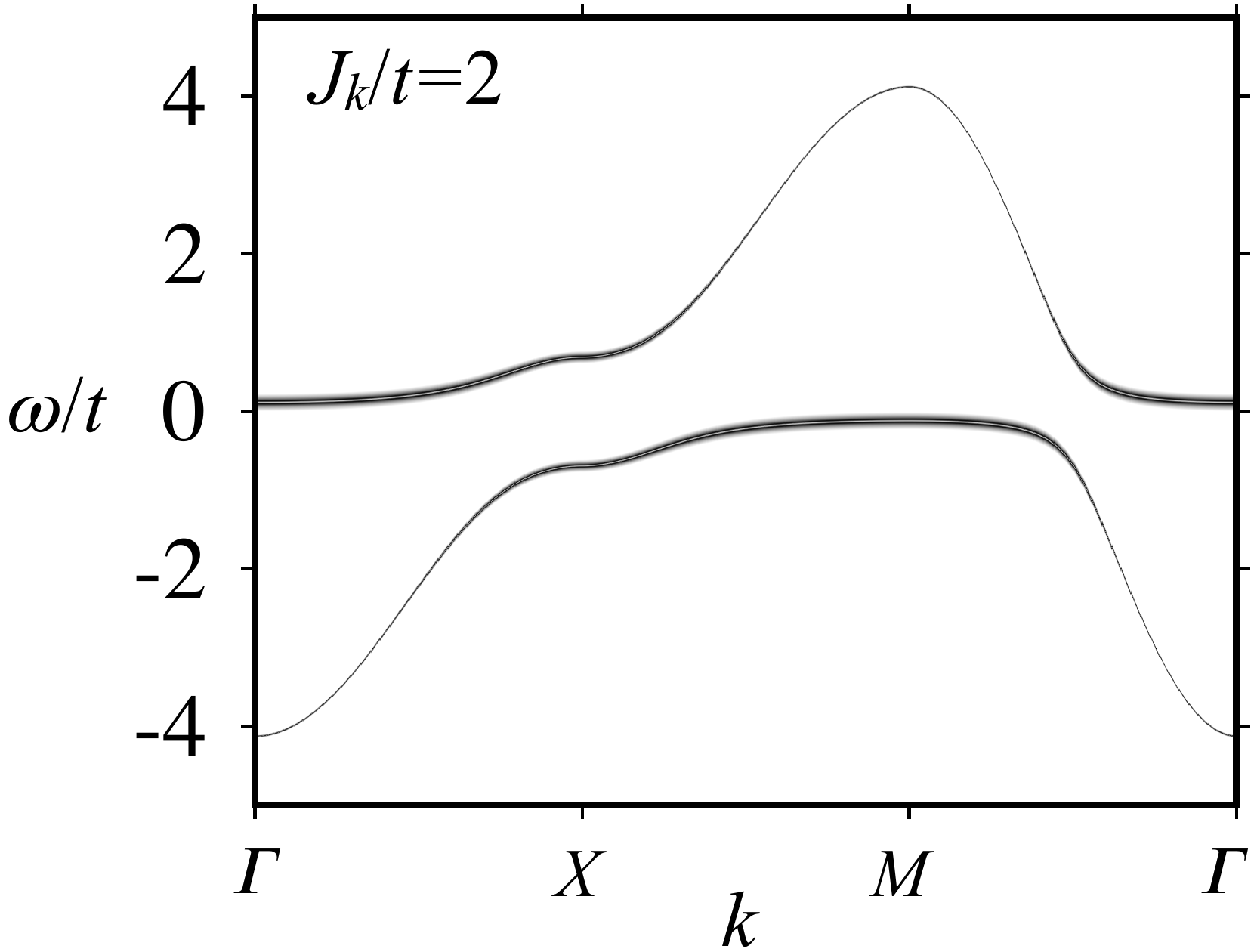}%
\caption{Spectral weight  $A_f(\k,\omega)$  as a function of momentum $\k$ and energy $\omega/t$  for $J_k/t=2$ as obtained in the large-$N$ limit.   The line  thickness reflects the $f$-character of the band. } 
\label{fspectral_LargeN_V} 
\end{figure}

Specifically, in the large-$N$ limit, the dispersion relation of the  $f$-electron  at 
 finite hybridization $V=\frac{2}{N}\big< \hat{D}^{\dagger}_{\i}  \big> =\frac{2}{N}\big< \hat{D}_{\i}  \big>$  is  given by    
\begin{eqnarray}
E_{\k,\pm}=\frac{1}{2} \Big(\epsilon_\k\pm\sqrt{\epsilon^2_\k+ J^2_kV^2}\Big)
\label{Enl_vs_k_sq}
\end{eqnarray}
with $\epsilon(\ve{k})=   -2 t\left(  \cos{k_x} +  \cos{k_y} \right) $.  The  $f$-electron spectral function
$A_f(\k,\omega) =-\frac{1}{\pi} ~\text{Im}  ~ G^{\text ret}_f(\k,\omega)$  reads:
 \begin{eqnarray}   
G^{\text ret}_f(\k,\omega)=\frac{|u_\k|^2}{\omega+i0^+-E^-_{\k}}+\frac{|v_\k|^2}{\omega+i0^+-E^+_{\k}}
\end{eqnarray}
with   coherence factors:
\begin{align} 
|u_\k|^2=\frac{1}{2} \left (1+ \frac{\epsilon_\k}{\sqrt{\epsilon^2_\k +(J_kV)^2}}\right ), \label{coh_uk} \\ 
|v_\k|^2=\frac{1}{2} \left (1- \frac{\epsilon_\k} {\sqrt{\epsilon^2_\k +(J_kV)^2}}\right ) \label{coh_vk}.  
\end{align}
Figure~\ref{fspectral_LargeN_V} plots $A_f(\k,\omega)$ as a function of energy and momentum. 
A dominant $f$-character of the low energy  heavy fermion band can be noticed by thick black lines. 

\subsection{Strong coupling  limit}\label{Strong_cpl}

An alternative  way  of   understanding the  composite  fermion  operator that  becomes  very transparent in the  strong coupling  limit, is in terms of   bond operators  between  conduction  electrons and   spins \cite{Jurecka01b,Feldbach02}.    We consider the states:  
\begin{eqnarray}
	   & & \hat{s}^{\dagger}_{\ve{i}}\big|0\big\rangle  =  \frac{1}{\sqrt{2}}   \left(\hat{c}^{\dagger}_{\ve{i}, \uparrow}  \hat{f}^{\dagger}_{\ve{i}, \downarrow}   
	    - \hat{c}^{\dagger}_{\ve{i}, \downarrow} \hat{f}^{\dagger}_{\ve{i}, \uparrow}   \right)\big|0\big\rangle     \nonumber  \\ 
	   & &\hat{t}^{\dagger}_{\ve{i},0}\big|0\big\rangle  =  \frac{1}{\sqrt{2}}   \left(\hat{c}^{\dagger}_{\ve{i}, \uparrow} \hat{f}^{\dagger}_{\ve{i}, \downarrow} 
	    +   \hat{c}^{\dagger}_{\ve{i}, \downarrow}   \hat{f}^{\dagger}_{\ve{i}, \uparrow}   \right)\big|0\big\rangle \nonumber    \\
	   & & \hat{t}^{\dagger}_{\ve{i},\sigma}\big|0\big\rangle  =  \hat{c}^{\dagger}_{\ve{i}, \sigma} \hat{f}^{\dagger}_{\ve{i}, \sigma}\big|0\big\rangle   \nonumber    \\
	  & & 
	     \hat{h}^{\dagger}_{\ve{i},\sigma}\big|0\big\rangle = \hat{f}^{\dagger}_{\ve{i},\sigma} \big|0\big\rangle   \nonumber    \\  & & 
	      \hat{d}^{\dagger}_{\ve{i},\sigma}\big|0\big\rangle =\hat{c}^{\dagger}_{\ve{i}, \uparrow} \hat{c}^{\dagger}_{\ve{i}, \downarrow}  \hat{f}^{\dagger}_{\ve{i},\sigma} \big|0\big\rangle.
\end{eqnarray}
Here, $\hat{s}^{\dagger}$ and  $\hat{t}^{\dagger}_{1,0,-1}$ denote a singlet and three triplet states (triplons)  with one conduction electron per site  and  $\hat{h}^{\dagger}_{\sigma}$ and  $\hat{d}^{\dagger}_{\sigma}$ denote holons  and doublons  of the conduction electrons.    We will  assume that holons and  doublons  (singlets and triplets) are independent fermionic (bosonic)  excitations such that    
$ \{ \hat{d}^{\#}_{\ve{i},\sigma}, \hat{d}^{\#'}_{\ve{j},\sigma'} \} = \{ \hat{h}^{\#}_{\ve{i},\sigma}, \hat{h}^{\#'}_{\ve{j},\sigma'} \}  = (1 -\delta_{\#,\#'}) \delta_{\ve{i},\ve{j} } \delta_{\sigma,\sigma'}  $,  
$ [ \hat{t}^{}_{\ve{i},m}, \hat{t}^{\dagger}_{\ve{j},m'} ] =  \delta_{\ve{i},\ve{j} } \delta_{m,m'}  $,  $ [ \hat{s}^{}_{\ve{i}}, \hat{s}^{\dagger}_{\ve{j}} ] =  \delta_{\ve{i},\ve{j} } $  and 
 $ [\hat{t}^{}_{\ve{i},m}, \hat{t}^{}_{\ve{j},m'} ] = [ \hat{s}^{}_{\ve{i}}, \hat{s}^{}_{\ve{j}} ] =0 $.   Here  $ \#  =  \cdot, \dagger $, and  the fermion and boson operators  commute.    
 To  suppress the  unphysical  states,   one then imposes the constraint: 
 \begin{eqnarray}
\hat{s}^\dagger_{\ve{i}}\hat{s}_{\ve{i}}+\sum_{m=1,0,-1}\hat{t}^\dagger_{\ve{i},m}\hat{t}_{\ve{i},m}+\sum_{\sigma=\uparrow,\downarrow}(\hat{h}^\dagger_{\ve{i},\sigma} \hat{h}_{\ve{i},\sigma}+\hat{d}^\dagger_{\ve{i},\sigma} \hat{d}_{\ve{i},\sigma})=1.\nonumber\\
\end{eqnarray}

In this representation the conduction electron and the  composite  fermion operators read:
\begin{eqnarray}
\label{c_bond.eq}
\hat{c}^\dagger_{\i,\sigma}=\frac{\sigma}{\sqrt{2}} (\hat{s}^\dagger_{\ve{i}}+\sigma \hat{t}^\dagger_{\ve{i},0}) \hat{h}_{\ve{i},-\sigma}+\hat{t}^\dagger_{\ve{i},\sigma} \hat{h}_{\ve{i},\sigma} \nonumber    \\-\frac{\hat{d}^\dagger_{\ve{i},\sigma}}{\sqrt{2}} (\hat{s}_{\ve{i}}-\sigma \hat{t}_{\ve{i},0}) +\sigma \hat{d}^\dagger_{\ve{i},-\sigma} \hat{t} _{\ve{i},-\sigma}
\end{eqnarray}
\begin{eqnarray}
\label{psi_bond.eq}
2\hat{\psi}^\dagger_{\i,\sigma}=-\frac{\sigma}{\sqrt{2}} (\hat{s}^\dagger_{\ve{i}}+\sigma \hat{t}^\dagger_{\ve{i},0}) \hat{h}_{\ve{i},-\sigma}+(\hat{t}^\dagger_{\ve{i},\sigma} +2\hat{t}^\dagger_{\ve{i},-\sigma}) \hat{h}_{\i,\sigma} \nonumber    \\-\frac{\hat{d}^\dagger_{\ve{i},\sigma}}{\sqrt{2}} (\hat{s}_{\ve{i}}-\sigma \hat{t}_{\ve{i},0}) -\sigma d^\dagger_{\ve{i},-\sigma} (\hat{t} _{\ve{i},-\sigma}+2\hat{t} _{\ve{i},\sigma}).\nonumber\\
\end{eqnarray}

As apparent  both the conduction   electron and  composite fermion creation  operators have very similar forms. In both cases  a holon (triplon or  singlet)  can be annihilated  to  generate a triplon or singlet (doublon).    We will now  show  that in the strong  coupling limit,  both the  composite fermion    and conduction electron  spectral functions share the same features   consisting of  valence and conduction bands. 
In the limit  $J_k\rightarrow  \infty $   triplons  can be neglected since in a given  fixed particle number Hilbert space the  triplon cost  is set by $J_k$.   Adopting this approximation, 
\begin{equation}
	2\hat{\psi}^\dagger_{\i,\sigma}=-\frac{\sigma}{\sqrt{2}} \hat{s}^\dagger_{\ve{i}} \hat{h}_{\ve{i},-\sigma}
	  -\frac{1}{\sqrt{2}}\hat{d}^\dagger_{\ve{i},\sigma} \hat{s}_{\ve{i}} 
\end{equation} 
and the Hamiltonian  reads: 
\begin{eqnarray}
  \hat{H}_{KLM}   = & &  -\frac{t}{2}   \sum_{\langle \ve{i},\ve{j}\rangle, \sigma}  \left(  
   \hat{s}^\dagger_{\ve{i}} \hat{h}^{}_{\ve{i},-\sigma}  \hat{h}^{\dagger}_{\ve{j},-\sigma}  \hat{s}^{}_{\ve{j}}  + \hat{d}^\dagger_{\ve{i},\sigma}  \hat{s}^{}_{\ve{i}}   \hat{s}^{\dagger}_{\ve{j}}   \hat{d}^{\phantom\dagger}_{\ve{j},\sigma}      + \text{H.c.} \right)      \nonumber  \\ 
       & & -\frac{3 J_k}{4}  \sum_{\ve{i}} \hat{s}^{\dagger}_{\ve{i}} \hat{s}^{\phantom\dagger}_{\ve{i}}.
 \end{eqnarray}
In the  above,   we have neglected   terms   such as
 $ \hat{d}^{\dagger}_{\ve{i},\sigma}  \hat{h}^{\dagger}_{\ve{j},-\sigma}   \hat{s}^{}_{\ve{i}} \hat{s}^{}_{\ve{j}} $ that    create  holon doublon excitations  since these  processes, in a given fixed  particle-number  Hilbert space,     involve  an  excitation  gap  of  $3J_k/2$.   At  $T=0$  and at half filling  where the ground state   corresponds to a product   state of  singlets, the   retarded  Green's function  reads: 
 \begin{eqnarray}
 	4 G^{\text ret}_{\psi}(\k,\omega)  & = &    \frac{1}{\omega - \left( \frac{3J_k}{4} +   \frac{\epsilon( \ve{k})}{2} \right)  + i 0^{+} } \nonumber  \\ 
	  & + & \frac{1}{\omega - \left( - \frac{3J_k}{4} +   \frac{\epsilon( \ve{k})}{2} \right)  + i 0^{+} }
 \end{eqnarray}
with $\epsilon(\ve{k})=   -2 t\left(  \cos{k_x} +  \cos{k_y} \right) $   the   conduction  electron dispersion relation.   The valence and conduction bands  are  separated  by  an indirect gap  since the maximal (minimal)  energy of the   valence (conduction) band is    at $\ve{k} =  (\pi,\pi) $  ($\ve{k} =  (0,0)$).      An  equivalent form can be  obtained for the conduction electron   spectral function.  The   obtained  dispersion relation   compares  with the  strong coupling expansion  presented in Ref.~\cite{Tsunetsugu97_rev}.

\subsection{Large-$S$ limit}\label{largeSthry}
We now consider  the large-$S$  limit.  Here  we systematically enhance the  dimension of the  representation of the SU(2)  group of the  impurity spins  such that the Hamiltonian is identical to that of Eq.~(\ref{model_ham})   but with 
\begin{equation}
	   \hat{\ve{S}}_{\i}^{2}  =  S(S+1)
\end{equation}  
for  half integer spins. 
 Since  we are  working on  bipartite lattices,  we foresee antiferromagnetic order and  choose the following  Holstein-Primakov representation of the spin algebra  on the  A 
 \begin{equation} 
 	\hat{S}^{z}_{\i} = S - \hat{b}^{\dagger}_{\i} \hat{b}^{\phantom\dagger}_{\i},  \,   \,  \hat{S}^{+}_{\i} =  \sqrt{ 2S -\hat{b}^{\dagger}_{\i} \hat{b}^{\phantom\dagger}_{\i}  }  \,  \hat{b}^{\phantom\dagger}_{\i}
 \end{equation}
 and   B  sub-lattices 
 \begin{equation}
\hat{S}^{z}_{\i} =  \hat{b}^{\dagger}_{\i} \hat{b}^{\phantom\dagger}_{\i} -S ,  \,   \,  \hat{S}^{+}_{\i} =    \hat{b}^{\dagger}_{\i}   \sqrt{ 2S -\hat{b}^{\dagger}_{\i} \hat{b}^{\phantom\dagger}_{\i}  } . 
 \end{equation}
 The N\'eel  state   corresponds to the vacuum of the boson operator  $ \hat{b}_{\i} $.  Hence in the large-$S$  limit  and in the N\'eel state we will assume that $ \langle \hat{b}^{\dagger}_{\i} \hat{b}^{\phantom\dagger}_{\i}   \rangle  << 2S  $  allowing one to expand   the square root. 
  With this  approximation,  and including the  RKKY interaction,  
  \begin{equation}
      J_{RKKY} \sum_{ \left< \i,\j  \right>}  \hat{\ve{S}}_{\ve{i}}  \cdot \hat{\ve{S}}_{\ve{j}} 
  \end{equation}
  to the KLM, we obtain:
 \begin{eqnarray}
        \hat{H}  = & &  \sum_{ \k,\sigma} \epsilon_{\k}{\hat c}^{\dagger}_{\k,\sigma}  \hat c_{\k,\sigma}
          +\frac{J_k}{2} S   \sum_{\i}   e^{i \ve{Q}\cdot \ve{i}}   \hat{\ve{c}}^{\dagger}_{\i}  \ve{\sigma}^z \hat{\ve{c}}^{}_{\i}  + \nonumber \\
          + & &  \frac{J_{RKKY}S} {2} \sum_{\i,  \ve{\delta}}   \big(  2 \hat{b}^{\dagger}_{\i}  \hat{b}^{\phantom\dagger}_{\i} + \hat{b}^{\phantom\dagger}_{\i}  \hat{b}^{\phantom\dagger}_{\i+ \ve{\delta}} 
           +  \hat{b}^{\dagger}_{\i}  \hat{b}^{\dagger}_{\i+ \ve{\delta}}  \big)   \nonumber   \\  
           + & &    J_k \sqrt{ \frac{S}{2} }  \sum_{\i \in A }  \big(  \hat{c}^{\dagger}_{\i, \uparrow} \hat{c}^{\phantom\dagger}_{\i, \downarrow} \hat{b}_{\i}^{\dagger}  + \hc    \big)  +  \nonumber \\
           + & &    J_k \sqrt{ \frac{S}{2} }  \sum_{\i \in B }  \big(  \hat{c}^{\dagger}_{\i, \uparrow} \hat{c}^{\phantom\dagger}_{\i, \downarrow} \hat{b}_{\i}^{\phantom\dagger}  + \hc    \big)  +  \cdots. 
 \label{Ham_largeS}
 \end{eqnarray}
 In the above, the ellipsis   corresponds to terms in lower order in $S$, $\ve{\delta} $ runs over the nearest neighbors of a  given site,  and $ e^{i \ve{Q}\cdot \ve{i}}  $  takes the value  $1$ ($-1$)  on the A (B)   sub-lattice.  

We now turn our attention to the  composite fermion  operator.   Using the  Holstein-Primakov representation we  obtain:
\begin{eqnarray}	
	 \hat{\ve{\psi}}^{\dagger}_{\i}  =  & &     S\hat{\ve{c}}^{\dagger}_{\i} \sigma^{z} e^{i \ve{Q}\cdot \ve{i}}  +   \sqrt{\frac{S}{2}}  \hat{\ve{c}}^{\dagger}_{\i}    \frac{1 + e^{i \ve{Q}\cdot \ve{i}}  } {2}  
	 \big(   \sigma^{+}  \hat{b}^{\dagger}_{\i}  +  \sigma^{-}  \hat{b}^{\phantom\dagger}_{\i}   \big)   \nonumber \\ 
	 + & &  \sqrt{\frac{S}{2}}  \hat{\ve{c}}^{\dagger}_{\i}    \frac{1 - e^{i \ve{Q}\cdot \ve{i}}  } {2}  
	 \big(   \sigma^{+}  \hat{b}^{\phantom\dagger}_{\i}  +  \sigma^{-}  \hat{b}^{\dagger}_{\i}   \big)  + \cdots.
\end{eqnarray}
 In the above  the ellipsis  again  refers to lower order  terms in $S$ and  $ \sigma^{\pm}  = \sigma^{x} \pm  i \sigma^{y} $.    Retaining terms  only up to order $S$,    the Hamiltonian of Eq.~(\ref{Ham_largeS}) corresponds to  electrons subject to a static staggered magnetic field of  magnitude $J_k S$  in the $z$-direction as well as   spin waves.    The composite fermion operator reduces to  the  conduction electron operator  with a phase shift such  that 
 \begin{eqnarray}\label{ferro_phaseQ} 
\sum_{\sigma}  \big \langle   \hat{{\psi}}^{\phantom\dagger}_{\i,\sigma}(0)  \hat{{\psi}}^{\dagger}_{\j,\sigma}(\tau)    \big \rangle  = & & S^2 \sum_{\sigma}   \big \langle  \hat{c}^{\phantom\dagger}_{\i,\sigma}(0)  \hat{c}^{\dagger}_{\j,\sigma}(\tau)   \big \rangle  e^{i \ve{Q}\cdot ( \ve{j} - \ve{i})} \nonumber \\ 
      + & & \cdots .	
  \end{eqnarray}
 Hence in the  large-$S$  limit, the composite fermion  Green's  function should  correspond to the $c$-Green's function with a momentum 
shift of $ \ve{Q} $.  We  note that the above equation can  also be motivated  from Eq.~(\ref{Eq.bubble}) with 
$\langle  \hat{\ve{S}}_{i} \hat{\ve{S}}_{j}(\tau)    \rangle   \propto  S^2   e^{i \ve{Q}\cdot ( \ve{j} - \ve{i})}   $  as appropriate for  a long ranged antiferromagnetic state. 
\section{Quantum Monte Carlo}
\label{QMC_method} 
We consider the SU($N$) generalization of the Kondo lattice Hamiltonian given in Eq.~(\ref{model_ham}):
\begin{eqnarray}
\Hhat&=&\Hhat_t-\frac{J_k}{4N}\sum_{\ve{i}}  \Big\{ \big(\ve{\hat{c}}^{\dagger}_{\ve{i} }\ve{\hat{f}}^{}_{\ve{i} }+\hc\big)^2+ \big(i\ve{\hat{c}}^{\dagger}_{\ve{i} }\ve{\hat{f}}^{}_{\ve{i}}+\hc\big)^2 \Big\}\nonumber\\&&
+\frac{U}{N}\sum_{\ve{i}} \Big(\ve{\hat{f}}^\dagger_{\ve{i}} \ve{\hat{f}}_{\ve{i} }-\frac{1}{2}\Big)^2
\label{ham_sq}
\end{eqnarray}
where $\Hhat_t=-t \sum_{\langle \ve{i},\ve{j}\rangle} ( \ve{\hat{c}}^{\dagger}_{\ve{i} }\ve{\hat{c}}^{}_{\ve{j} } +\hc) $ and 
$\ve{\hat{c}}^{\dagger}_{\ve{i}}$ and $\ve{\hat{f}}^{\dagger}_{\ve{i}}$ are $N$ flavor fermion operators.   
Since   the Monte Carlo sampling is formulated in an   unconstrained  Hilbert  space,   
to impose the constraint Eq.~(\ref{constraint})  we have added above a Hubbard-$U$ interaction  acting on the $f$-electrons.  Importantly,  the Hubbard interaction
commutes with the  Hamiltonian such that the  constraint  is very efficiently  implemented.

Next, using the Hubbard-Stratonovich  transformation the partition function can be written as:
\begin{eqnarray}
Z~\equiv~\int \mathcal{D}\{ z , \mathcal{\lambda} \} ~e^{- N \mathcal{S}\{ z ,\mathcal{\lambda}\}}
\end{eqnarray}
with the  action
\begin{eqnarray}
\mathcal{S}\{z,\lambda\}&=&-\ln \Big[~\mbox{Tr} ~ \mathcal{T} e^{-\int^\beta_0 d\tau~ \Hhat \{z ,\lambda\}}\Big] \nonumber\\&&+\int^\beta_0 d \tau \sum_{\ve{i}} \Big\{\frac{J_k}{4}| z( \ve{i},\tau)|^2 +\frac{ U}{4} |\lambda( \ve{i},\tau)|^2\Big\}\nonumber \\
 \end{eqnarray}
and  time dependent Hamiltonian
\begin{eqnarray}
 \Hhat \{z,\lambda\}&=&\Hhat_t+ \sum_{\ve{i}} \Big\{-\frac{J_k}{2}\big(z( \ve{i},\tau) \ve{\hat{c}}^{\dagger}_{\ve{i} }\ve{\hat{f}}^{}_{\ve{i} }+\hc\big)\nonumber\\&&-i U\lambda(\ve{i},\tau) \Big(\ve{\hat{f}}^\dagger_{\ve{i}} \ve{\hat{f}}_{\ve{i} }-\frac{1}{2}\Big) \Big\}.
\end{eqnarray}
In the  above scalar   field  $\mathcal{\lambda}( \ve{i},\tau)$ enforces the constraint and $z( \ve{i},\tau)$  is a space and time dependent complex bond field. For a particle-hole symmetric band the imaginary part of the action takes the value  $ n \pi$ with $n$ an integer resulting in no negative sign problem for even $N$~\cite{Assaad99a,Capponi00,Raczkowski20}. Note that the on-site Hubbard term commutes with the Hamiltonian. Hence  for a given $U$ and inverse temperature $\beta$    the unphysical even-parity states are  suppressed by a factor $e^{-\beta U/N}$. The  choice  $\beta U/N (\ge 10)$  allows  restriction of the 
Hilbert space to  the odd parity sector  within our error bars.

We have used both the finite temperature~\cite{Blankenbecler81,White89} as well as the zero temperature auxiliary field QMC 
methods~\cite{Assaad99a,Assaad08_rev,ALF_v1}. For the SU(2) invariant KLM we have mainly used the finite temperature algorithm. 
In that case, it is possible to perform simulations at low enough temperatures  to address the interplay between strong antiferromagnetic spin 
correlations and Kondo screening. Given that the RKKY scale varies as $1/N$ in the SU($N$) generalization of the KLM, 
we have opted for $N>2$  for a projective QMC technique based on the imaginary-time evolution of a trial wave function 
$| \Psi_\text{T}\rangle$, with $ \langle\Psi_\text{T}  |\Psi_0 \rangle  \neq 0 $,  to the ground state $|\Psi_0 \rangle$:             
\begin{equation}
  \frac{ \langle  \Psi_0 | \hat{O} |  \Psi_0 \rangle  }{ \langle  \Psi_0  |  \Psi_0 \rangle  }  =  
  \lim_{\Theta \rightarrow \infty} 
  \frac{ \langle  \Psi_\text{T} | e^{-\Theta \hat{\mathcal{H}}  } \hat{O} e^{-\Theta \hat{\mathcal{H}}   } |  \Psi_\text{T}\rangle  }
  { \langle  \Psi_\text{T}  | e^{-2\Theta \hat{\mathcal{H}} }  |  \Psi_\text{T} \rangle  }
\end{equation}
where the  projection parameter $\Theta$  is chosen to be sufficiently large (up to $2\Theta t=400$ for our largest $N=8$) 
to reach the antiferromagnetic ground state even at a small value of $J_k/t=0.4$.  
Finally, taking into account the phase diagram presented in Refs.~\cite{Assaad99a,Capponi00,Raczkowski20}  we have considered values of 
$J_k/t$ representative of  both the magnetically  ordered RKKY and disordered Kondo screened phases.  

The SU($N$) KLM is one of the   standard Hamiltonians  implemented  in the  ALF2.0 library~\cite{ALF_v2}.   We  have used this library to  produce our  results and  
refer the reader to Ref.~\cite{ALF_v2}  for further details of the implementation.  We also note that  the  measurement of the composite fermion time displaced 
correlation functions is   implemented in ALF2.0   so that no add-ons  to the library are required to  reproduce the data presented in this paper.

\section{Quantum Monte Carlo results}\label{QMC_results}
Our main goal is to compute the momentum resolved composite fermion spectral function defined in Eqs.~(\ref{Afkomegaeq})~-~(\ref{Gfkomegaeq}). 
The motivation of doing so stems from the fact that the low energy heavy fermion band has predominantly $f$-character and thus the composite 
fermion Eq.~(\ref{composite_fermion}) may provide a better possibility to resolve heavy quasiparticles as compared with the spectral function 
on conduction electrons,  
$A_c(\k,\omega) =-\frac{1}{\pi} ~\text{Im}  ~ G^{\text ret}_c(\k,\omega)$, where 
 \begin{eqnarray}   
 G^{\text ret}_c(\k,\omega)=-i \int^{\infty}_0 d t e^{i \omega t} \sum_{\sigma} \big \langle \big\{ {\hat c}_{\k,\sigma}(t), {\hat c}^\dagger_{\k,\sigma}(0) \big\} \big\rangle.\nonumber \\
 \end{eqnarray}
To extract the real frequency data from  the imaginary time data of QMC we have used the stochastic analytical continuation 
algorithm~\cite{KBeach2004}  of the  ALF2.0 \cite{ALF_v2}  library. In the following we present and discuss  our results considering separately antiferromagnetic 
$J_k/t>0$ and ferromagnetic $J_k/t<0$ exchange couplings.

\subsection{Antiferromagnetic Kondo  lattice}

As proposed by Doniach~\cite{Doniach1977}, the antiferromagnetic Kondo coupling in the KLM leads to 
two energy scales set by Kondo and RKKY exchange interactions. The Kondo scale is given by a single impurity Kondo scale 
$T_k\sim We^{-W/J_k }$, where $W$ denotes the  bandwidth of the conduction electrons.  
The RKKY scale is given by $J_{RKKY}(\ve{q})=  J^2_k \chi^{c}(\ve{q},\omega =  0)$ where $\chi^{c}$ is the spin susceptibility of the conduction 
electrons. 
Depending on the magnitude of the exchange coupling, the physics is dominated 
by one of these two energy scales. For large Kondo couplings, the Kondo effect is the dominant effect and stabilizes 
the spin-gapped Kondo singlet phase. For small Kondo couplings, the RKKY interaction is the largest scale and 
thus magnetic order of local moments occurs. This competition leads to a quantum phase transition which for the SU(2) KLM on the 
square lattice is shown to arise at the critical point $J^c_k/t\simeq 1.45$~\cite{Assaad99a,Capponi00}.  The location of the magnetic transition 
shifts with increasing number of flavors $N$ in the SU($N$) generalization of the KLM towards smaller values of 
$J_k/t$~\cite{Raczkowski20}.  
Keeping in mind these two competing energy scales, we proceed to discuss their signature in the momentum resolved composite fermion 
spectral function $A_\psi(\k,\omega)$. By comparing the latter with the single particle spectrum of conduction electrons $A_c(\k,\omega)$ 
we will directly show the advantage of usage of $A_\psi(\k,\omega)$  in detecting the existence of heavy quasiparticles especially 
in the weak coupling region of the phase diagram.

\subsubsection {SU(2)  Kondo lattice model}

\begin{figure}[t]
\centering
\includegraphics[width=0.49\textwidth]{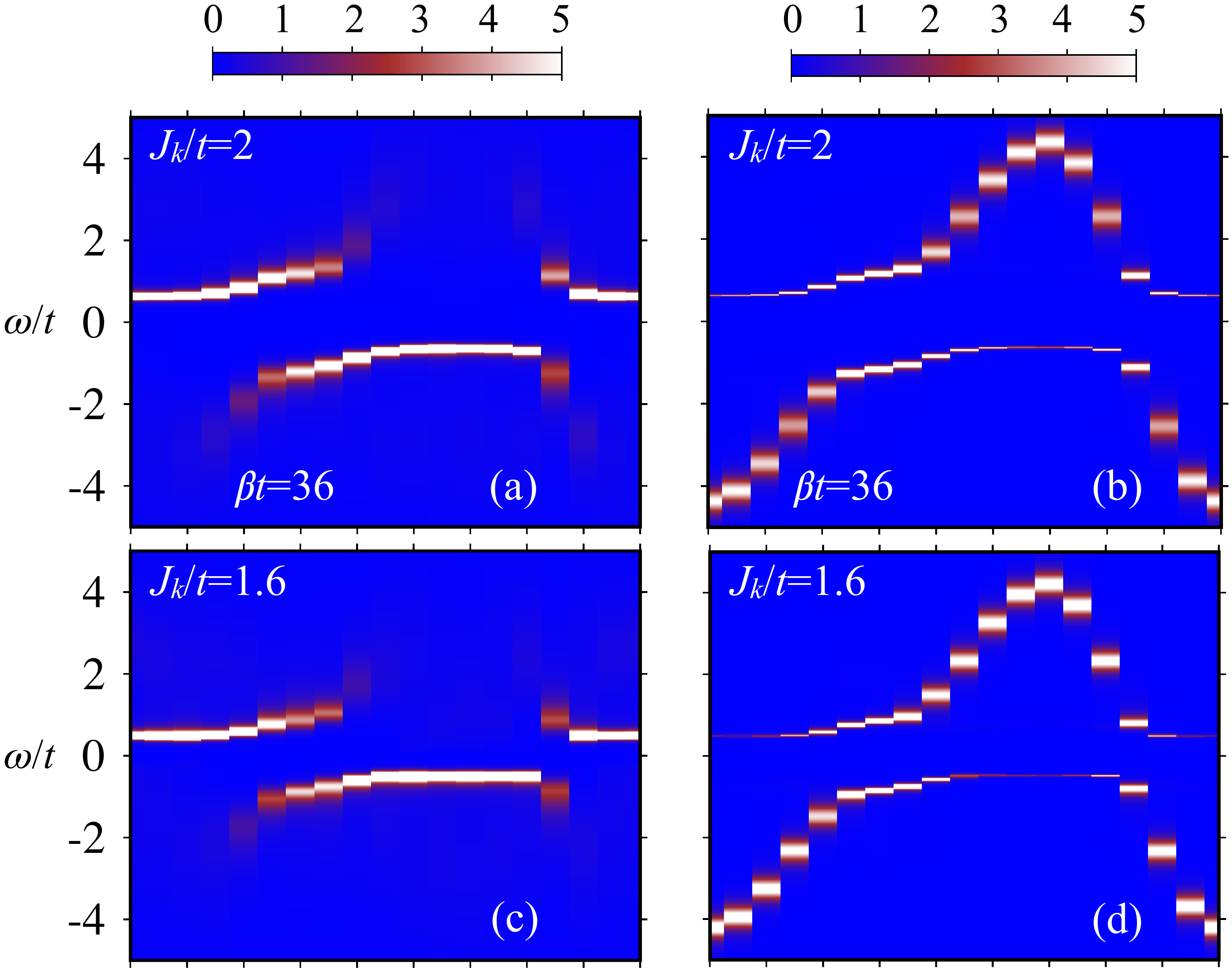}\\
\includegraphics[width=0.49\textwidth]{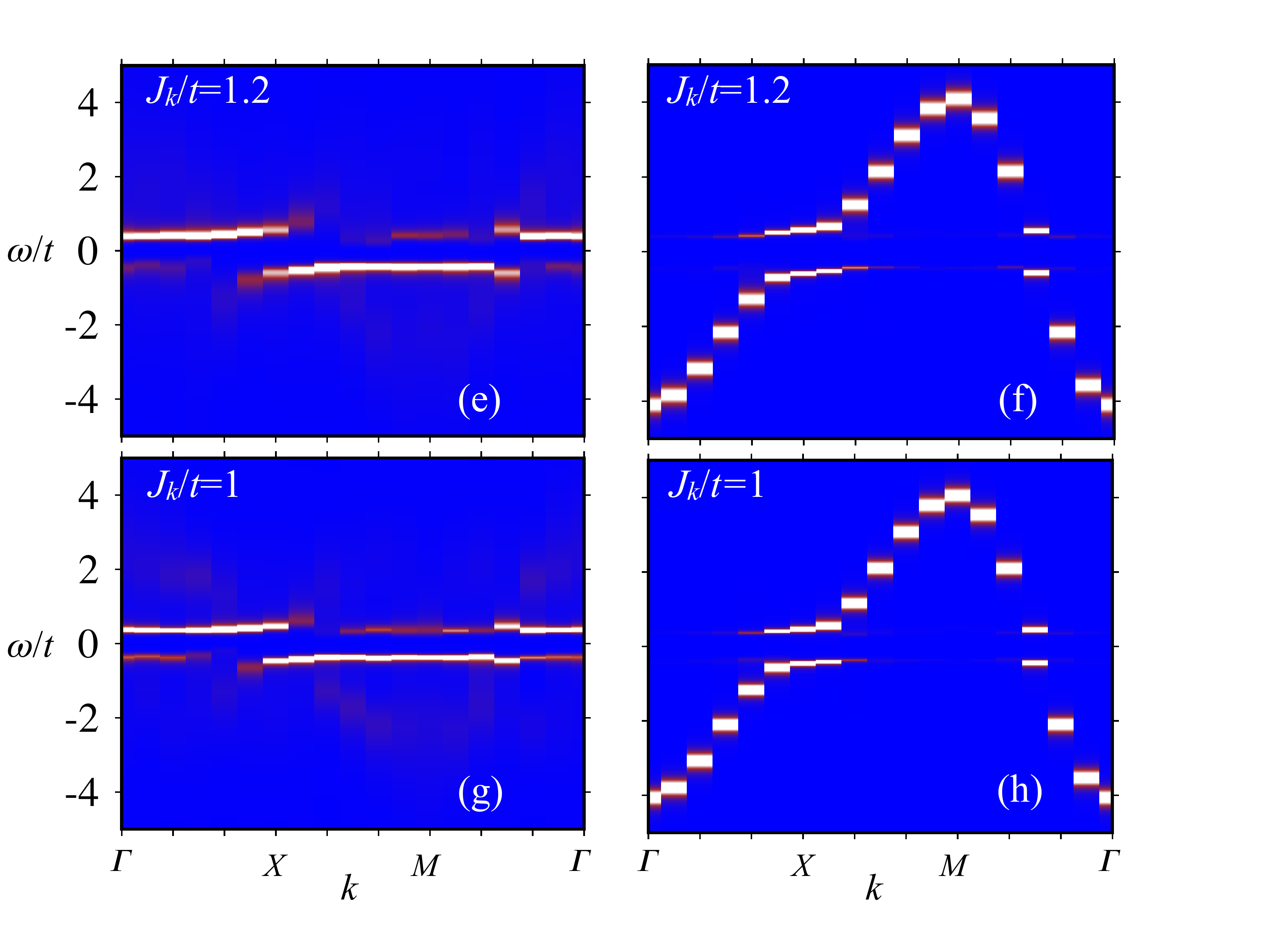} 
\caption{Composite fermion $A_\psi(\k,\omega)$ (left) and conduction electron  $A_c(\k,\omega)$ (right)  spectral functions 
in the $12\times 12$ KLM at $\beta t=36$ with: 
(a,b) $J_k/t=2$; (c,d) $J_k/t=1.6$; (e,f) $J_k/t=1.2$, and (g,h) $J_k/t=1$.
}
\label{Spectralcf_sq_L12_Beta36} 
\end{figure}

We start with finite but low temperature $T=t/36$ data of the SU(2) invariant KLM in the Kondo-screened phase.  
The Kondo-screened phase is adiabatically connected to the strong coupling limit, where each spin binds with a conduction 
electron into a spin singlet. As discussed in Sec.~\ref{Strong_cpl}, the composite fermion spectral function  $A_\psi(\k,\omega)$ 
should show a well defined quasiparticle band.  Figures~\ref{Spectralcf_sq_L12_Beta36}(a) with  $J_k/t=2$ 
and ~\ref{Spectralcf_sq_L12_Beta36}(c) with  $J_k/t=1.6$ confirm this point of view.   
Furthermore, although no actual symmetry breaking occurs in the QMC 
simulations, the low energy excitation spectrum is 
consistent with that found in a simple large-$N$ picture in which the hybridization gap opens up in the 
presence of a finite hybridization parameter $V$, see Fig.~\ref{fspectral_LargeN_V}.  
Given the sum rule $|u_\k|^2+|v_\k|^2=1$ of mean-field coherence factors defined in Eqs.~(\ref{coh_uk}) and (\ref{coh_vk}), 
the low energy part of the spectrum with the dominant $f$-character is poorly  represented in the large-$N$ conduction 
electron spectral function  $A_c(\k,\omega)$.  As is apparent in Figs.~\ref{Spectralcf_sq_L12_Beta36}(b) and 
\ref{Spectralcf_sq_L12_Beta36}(d) the same holds for the QMC data: the intensity of a weakly dispersive heavy fermion 
band in $A_c(\k,\omega)$ quickly drops upon approaching the $M$ point while being much more pronounced in the composite fermion spectral function  
$A_\psi(\k,\omega)$.   

\begin{figure}[t]
\centering
\includegraphics[width=0.46\textwidth]{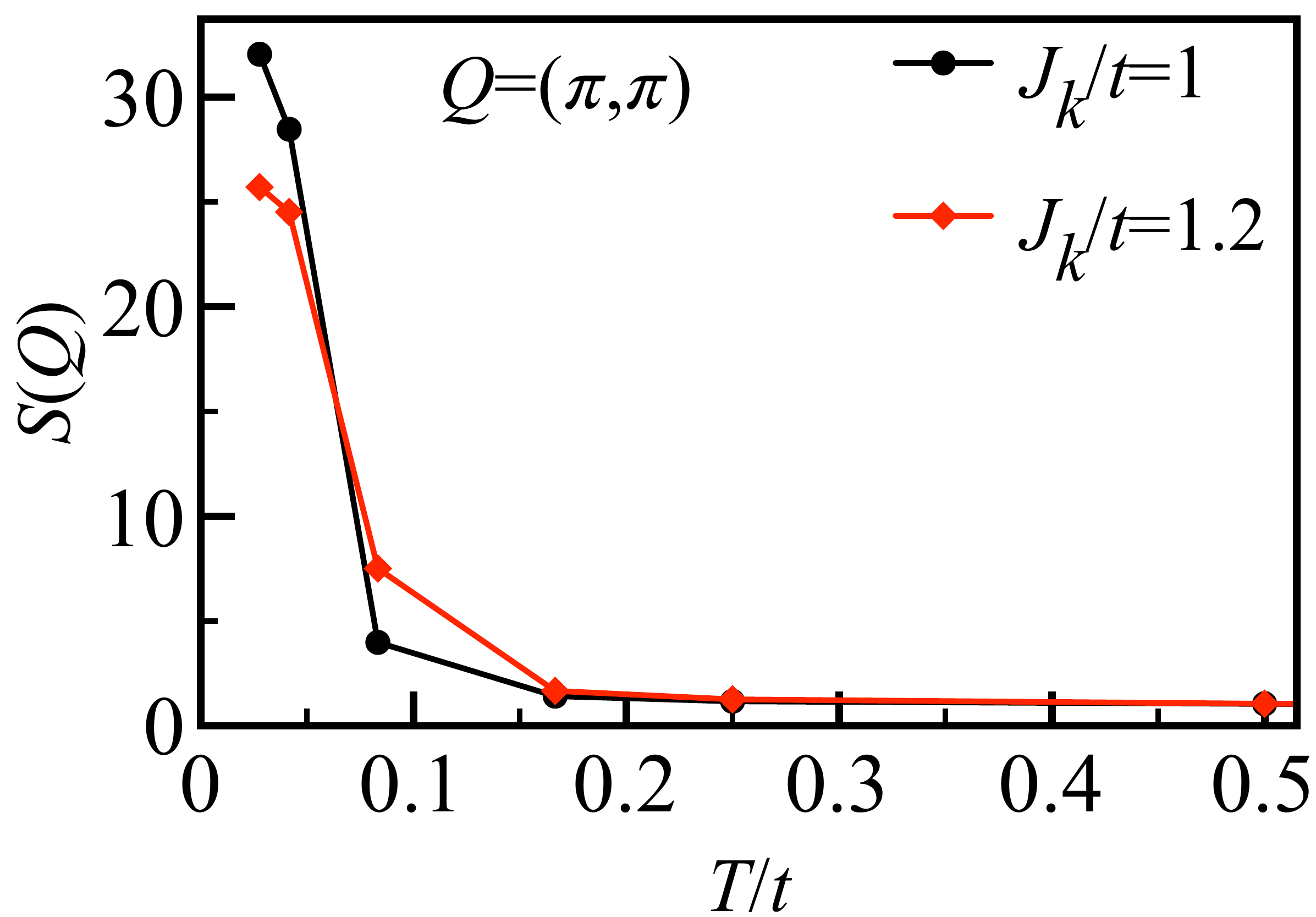} 
\caption{Static spin structure factor $S(\Q)$ at $\Q=(\pi,\pi)$ for the localized spins as a function of temperature $T/t$.
}  
\label{SQf_vs_T_SKLM_AFM} 
\end{figure}

To illustrate further a superior quality of the composite fermion as a tool in tracing heavy fermon excitations, 
we plot in Figs.~\ref{Spectralcf_sq_L12_Beta36}(e) and \ref{Spectralcf_sq_L12_Beta36}(g) its spectra in the RKKY dominated 
part of the phase diagram with $J_k/t=1.2$ and $J_k/t=1$. To make sure that the used temperature $T=t/36$ is low enough 
to access the interplay between Kondo screening and the RKKY interaction, we plot in Fig.~\ref{SQf_vs_T_SKLM_AFM}   
the corresponding  temperature dependence of a static spin structure factor
\begin{equation}
S(\Q)=\frac{1}{N_u} \sum_{\i,\j} e^{-i\Q(\i-\j)} \langle \hat{S}^z_\i\hat{S}^z_\j\rangle
\end{equation}
at the antiferromagnetic wave vector $\Q=(\pi,\pi)$. As can be seen, the onset of antiferromagnetic fluctuations begins 
already around $T=t/12$. Nevertheless, we easily note in $A_\psi(\k,\omega)$ the continued existence of the flat heavy fermion 
band around the $M$, in striking contrast with its barely visible fingerprint in the $c$-electron spectra, 
see Figs.~\ref{Spectralcf_sq_L12_Beta36}(f) and \ref{Spectralcf_sq_L12_Beta36}(h).

A closer inspection of $A_{\psi} (\k,\omega)$ reveals additional low energy band features located near the $\Gamma$ ($M$) 
momentum in the lower $\omega/t<0$  (upper $\omega/t>0$)  part of the spectrum, respectively. These shadow features emerge from the 
scattering of the heavy quasiparticle off the magnetic fluctuations with the wave vector $\Q=(\pi,\pi)$. 
Another consequence of strong antiferromagnetic spin correlations seen in $A_{\psi} (\k,\omega)$ is a faint image of
the $c$-electron band with a momentum shift of $\ve{Q}$, see Fig. \ref{Spectralcf_sq_L12_Beta36}(g). 
The emergence of this feature becomes clear by considering a simplified form of the composite fermion Green's function 
in Eq.~(\ref{ferro_phaseQ}) valid in the large-$S$ limit. 

Altogether, our finite-$T$ spectral data point towards the coexistence of Kondo screening and antiferromagnetism 
also in the broken spin symmetry phase that occurs at $T=0$.  We will confirm using the projective QMC method in Sec.~\ref{SUN}  
that this conjecture remains valid down to our smallest considered value $J_k/t=0.4$.

\begin{figure}[t]
\centering
\includegraphics[width=0.49\textwidth] {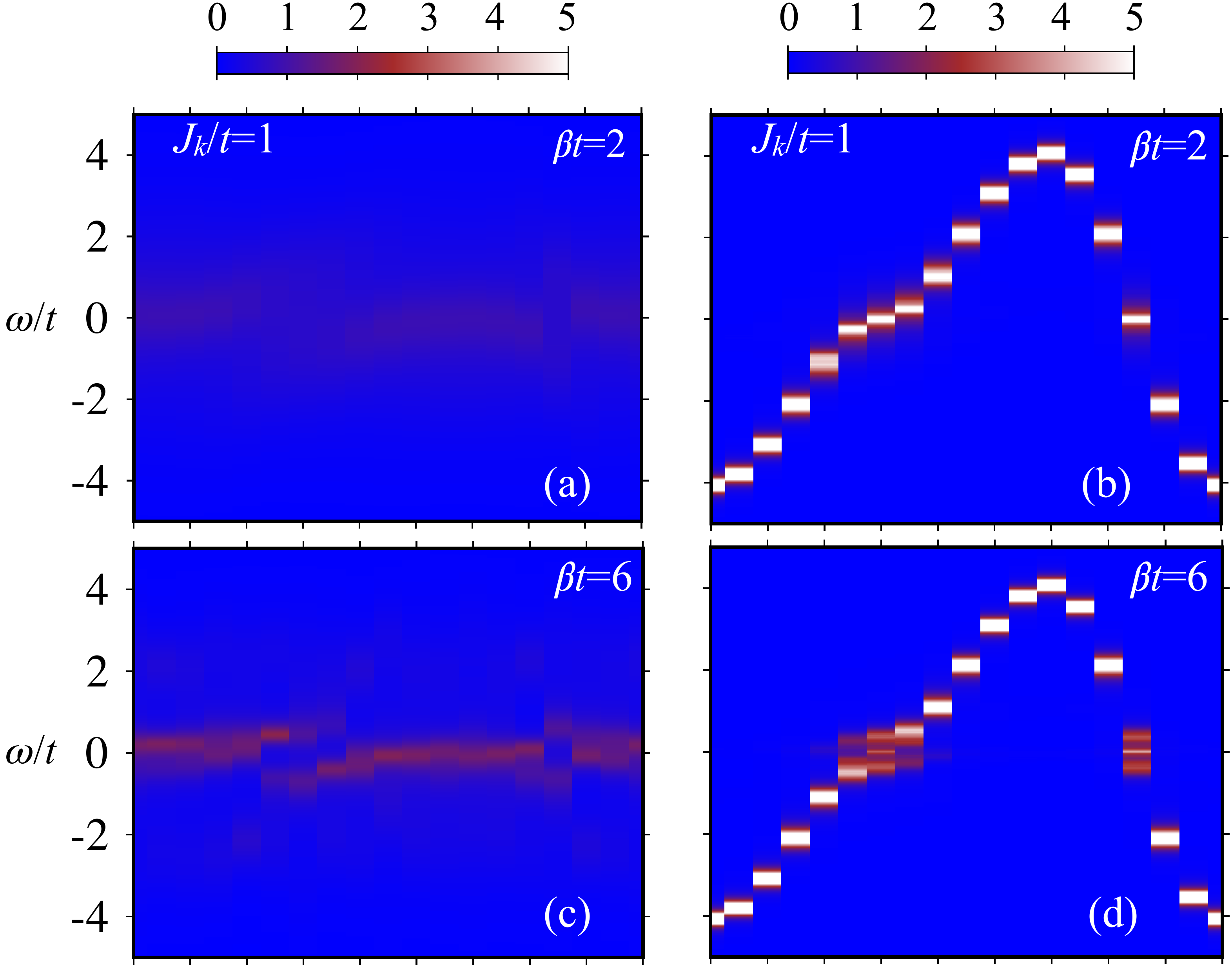}\\
\includegraphics[width=0.49\textwidth] {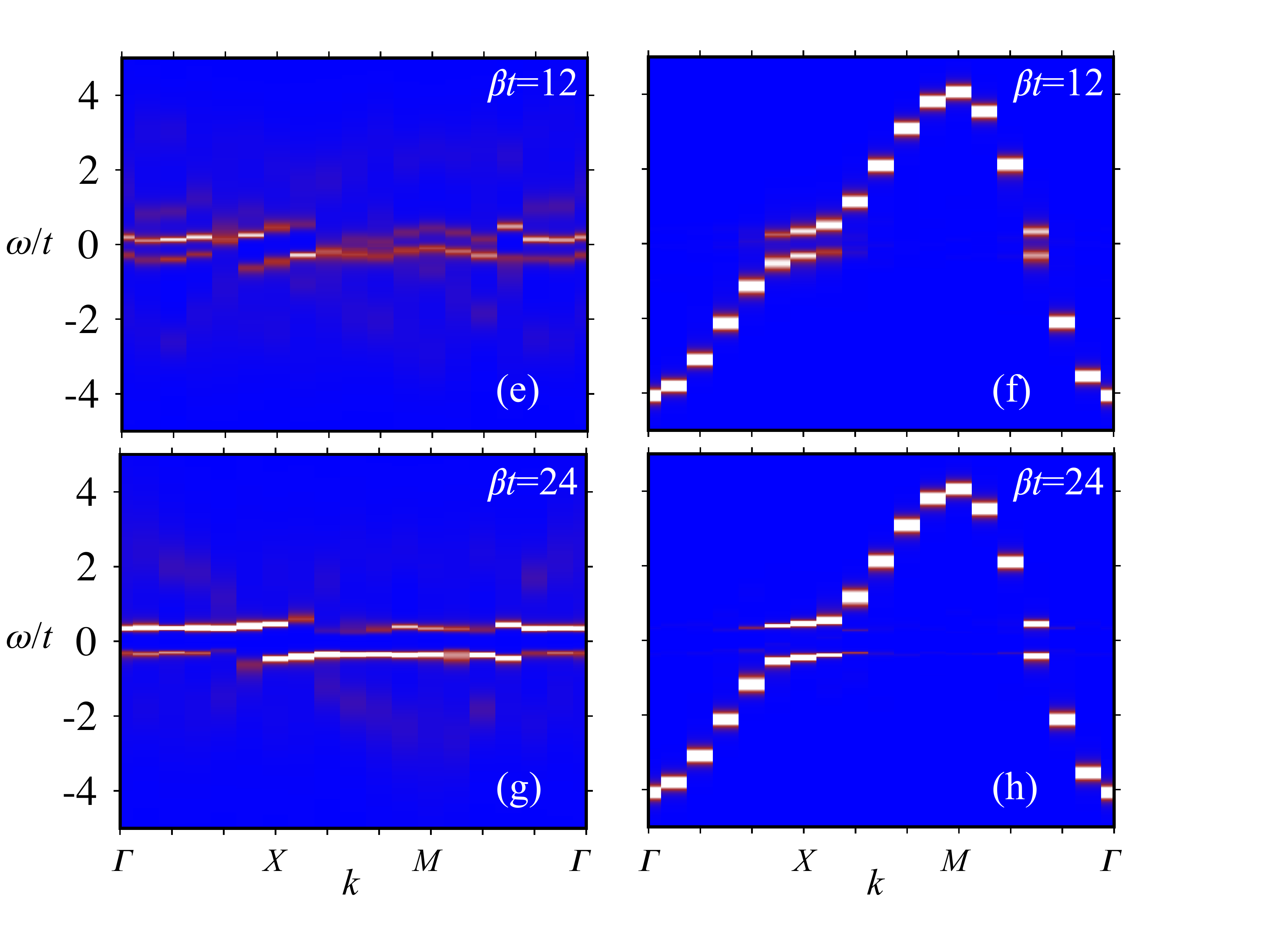}
\caption{Emergence of the heavy fermion band structure as seen in $A_\psi(\k,\omega)$ (left)  and $A_c(\k,\omega)$ (right) 
upon decreasing temperature in the $12\times 12$ KLM  at $J_k/t=1$: 
(a,b) $\beta t=2$; (c,d) $\beta t=6$; (e,f) $\beta t =12$, and (g,h) $\beta t=24$.
}  
\label{Jk1p_Spectralcf_sq_L12_Beta_dp} 
\end{figure}

\begin{figure}[t]
\centering
\includegraphics[width=0.46\textwidth]{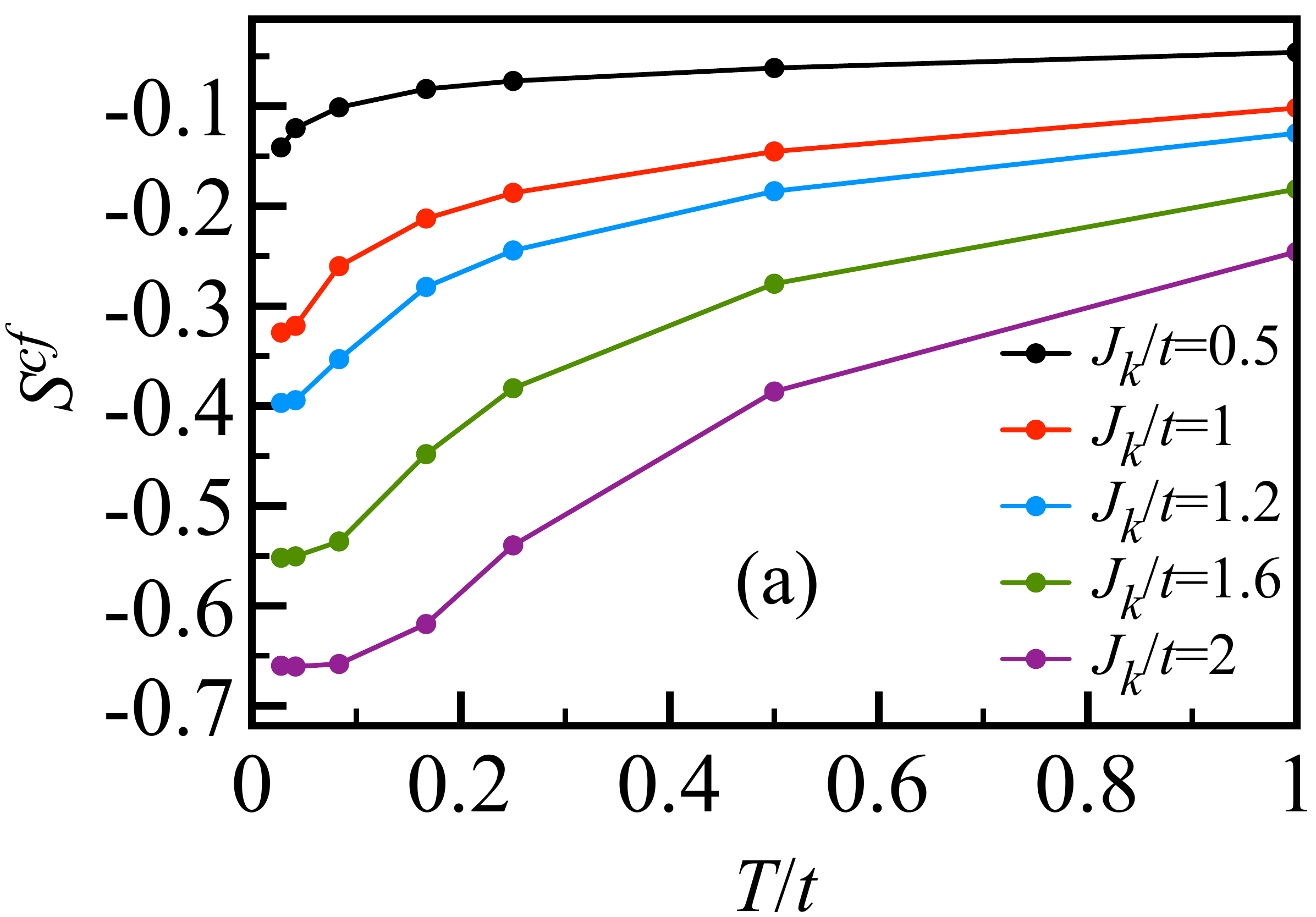}\\
\includegraphics[width=0.46\textwidth] {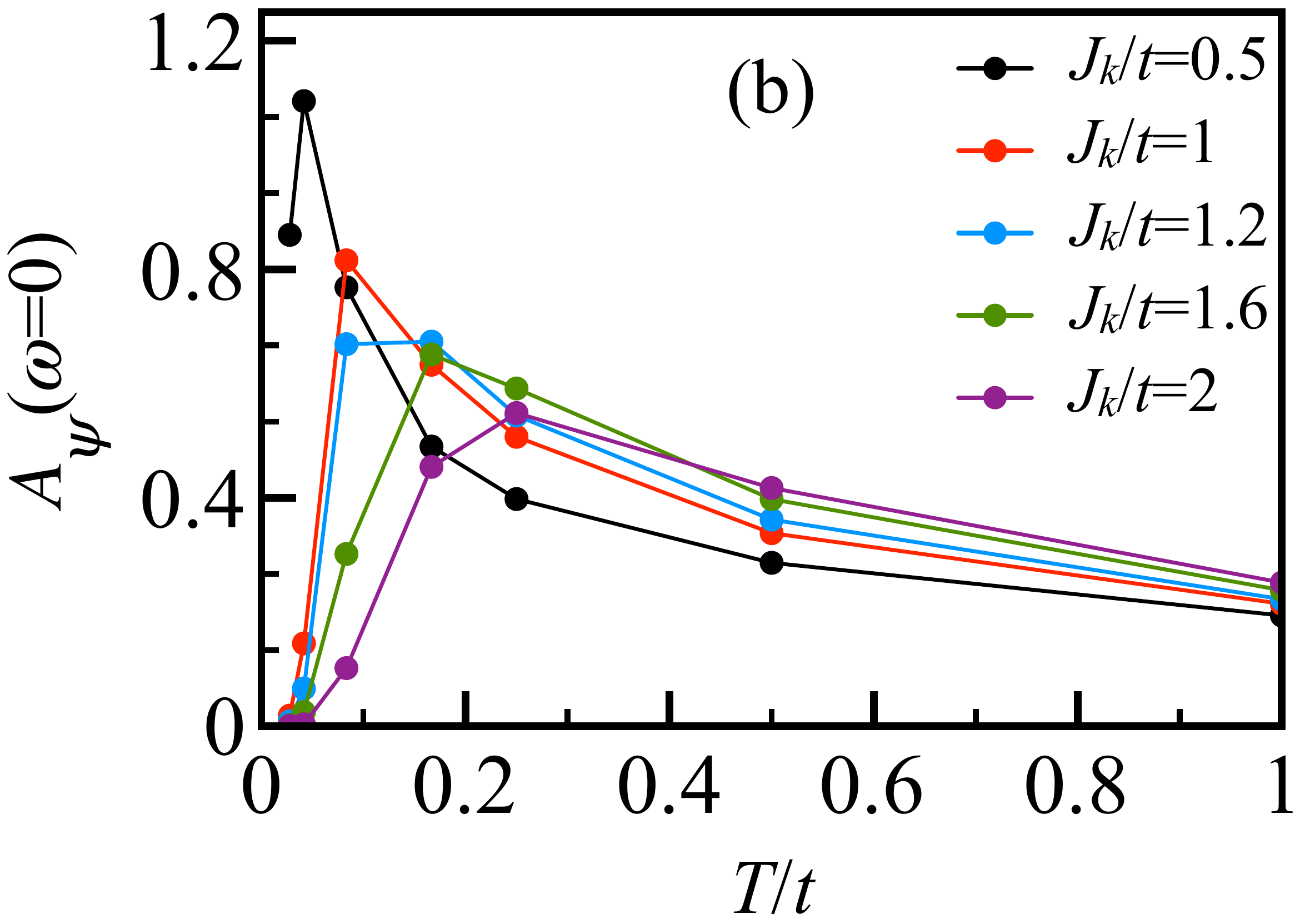}
\caption{(a) Local spin-spin correlation function $S^{cf}=\tfrac{1}{N_u}\sum_{\i}\langle\hat{\ve S}^c_{\i} \cdot \hat{\ve S}^{}_{\i}  \rangle$ and 
(b) estimate of the local density of states at the Fermi level 
$A_{\psi}(\omega=0) \approx \tfrac{1}{\pi}\tfrac{\beta}{N_u} \sum_{\k} G_{\psi}(\k,\tau=\beta/2)$  
as a function of temperature $T/t$ and for various Kondo couplings $J_k/t$.
The formation of local singlets seen as a decrease in $S^{cf}$ induces a depletion of the spectral weight  and opens a gap 
in $A_{\psi}(\omega=0)$ in the low-$T$ limit.} 
\label{dFbydJk_vs_T_SKLM} 
\end{figure}

We turn now to the discussion of how the electronic states in $A_\psi (\k,\omega)$ emerge and rearrange on passing  through 
progressively lower energy scales upon cooling the system. For concreteness, let us focus on the $J_k/t=1$ case illustrated in 
Fig.~\ref{Jk1p_Spectralcf_sq_L12_Beta_dp}. In the high temperature limit, it is legitimate to expect that the local moments do not 
interact with the spin degrees of freedom of the conduction electrons and are fully disordered.  Consequently, 
$A_\psi (\k,\omega)$ should have a $\ve{k}$-independent form given by the density of states of the conduction electrons, 
see Eq.~(\ref{Apsieq_kinpndt}). This is precisely observed in Fig.~\ref{Jk1p_Spectralcf_sq_L12_Beta_dp}(a) at $T=t/2$: 
a faint featureless cloud of the composite spectral weight is discernible only in a narrow window around the Fermi level 
reflecting the saddle point in $A_c(\k,\omega)$. As shown in Fig.~\ref{Jk1p_Spectralcf_sq_L12_Beta_dp}(b),  
the latter approaches that of the tight-binding model with the van Hove singularity in the density of states 
at $\omega=0$ yielding in turn the strongest signal in $A_\psi (\k,\omega\simeq 0)$. 

Upon lowering $T$, the screening of magnetic impurities becomes progressively important as signaled by a decreasing behavior of the 
local spin-spin correlation function 
$S^{cf}=\tfrac{1}{N_u}\sum_{\i}\langle\hat{\ve S}^c_{\i} \cdot \hat{\ve S}^{}_{\i}  \rangle$, see Fig.~\ref{dFbydJk_vs_T_SKLM}(a). 
The formation of bound states between conduction electrons and the $f$-spins gives rise to quasiparticle poles in the composite fermion Green's function.
As a result, the hybridized band structure becomes apparent in $A_\psi(\k,\omega)$, see Figs.~\ref{Jk1p_Spectralcf_sq_L12_Beta_dp}(c) 
and ~\ref{Jk1p_Spectralcf_sq_L12_Beta_dp}(e), with a concomitant suppression of the conduction electron spectral weight at the Fermi level.   
Finally, around $T=t/24$, $S^{cf}$ begins to saturate and thus the sum rule for the composite fermion spectral weight 
in Eq.~(\ref{EQ.sum_rule}) attains its maximum. As seen in Fig.~\ref{Jk1p_Spectralcf_sq_L12_Beta_dp}(g), $A_\psi (\k,\omega)$ is exhausted 
mainly by the intense heavy fermion band while the rest of the weight forms the shadow features. 

\begin{figure}[t] 
\centering
\includegraphics[width=0.45\textwidth] {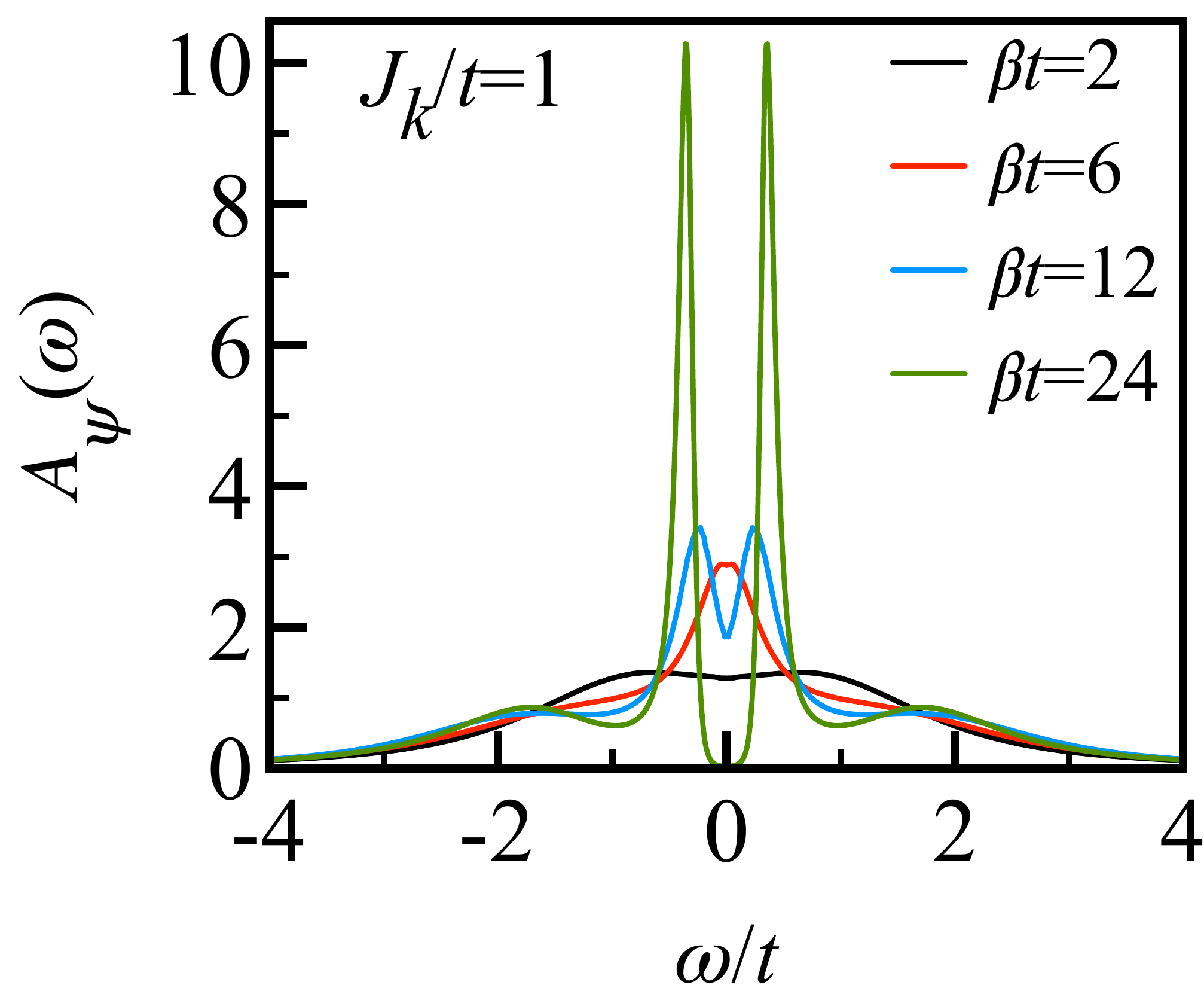}\\
\caption {Temperature evolution of the momentum integrated composite fermion spectral function 
	$A_{\psi}(\omega)=\tfrac{1}{N_u}\sum_{\k}A_\psi (\k,\omega)$ at $J_k/t=1$. }
\label{Aomega_vs_omega_sq}
\end{figure}

We conclude this section with pointing out that the crossover from free to screened magnetic impurities at low $T$ 
can be equally resolved in local density of states at the Fermi level $A_{\psi}(\omega=0)$. 
This quantity is directly related to the zero-bias differential conductance  $dI_{(V=0)}/dV$ in STM experiments. 
We compute it using an approximate form
\begin{equation}
A_{\psi}(\omega=0) \approx \frac{1}{\pi}\frac{\beta}{N_u} \sum_{\k} G_{\psi}(\k,\tau=\beta/2)
\label{LDOS}
\end{equation}
and plot its temperature dependence in Fig.~\ref{dFbydJk_vs_T_SKLM}(b). As is apparent, the formation of local singlets 
induces a depletion of the spectral weight  and opens a gap in $A_{\psi}(\omega=0)$ in the low-$T$ limit.
We confirm this in Fig.~\ref{Aomega_vs_omega_sq} which plots the temperature evolution of a momentum integrated composite fermion spectral 
function $A_{\psi}(\omega)=\tfrac{1}{N_u}\sum_{\k}A_\psi (\k,\omega)$ obtained with the analytical continuation of the 
imaginary time QMC  data. By lowering temperature one observes first an enhancement of the weight at the Fermi level followed, below $T=t/6$, 
by its transfer to  symmetrically developed about $\omega=0$  finite frequency peaks. 
In combination with the momentum resolved data in Fig.~\ref{Jk1p_Spectralcf_sq_L12_Beta_dp}(g), the sharp peaks at the lowest $T=t/24$ can be identified 
as upper and lower heavy fermion bands.

\subsubsection{SU($N$) Kondo lattice model}\label{SUN}

\begin{figure}[t]
\centering
\includegraphics[width=0.49\textwidth]{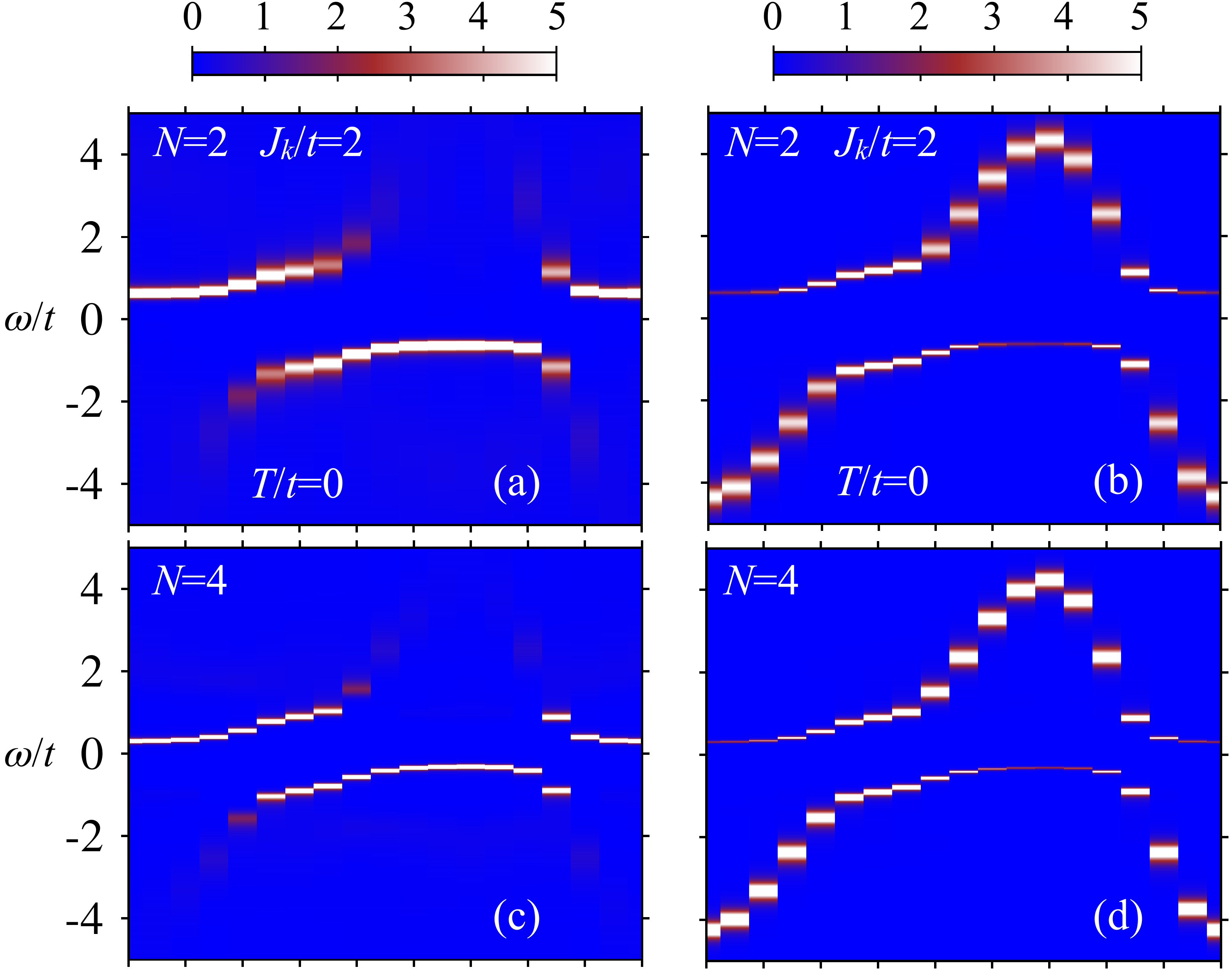}\\
\includegraphics[width=0.49\textwidth]{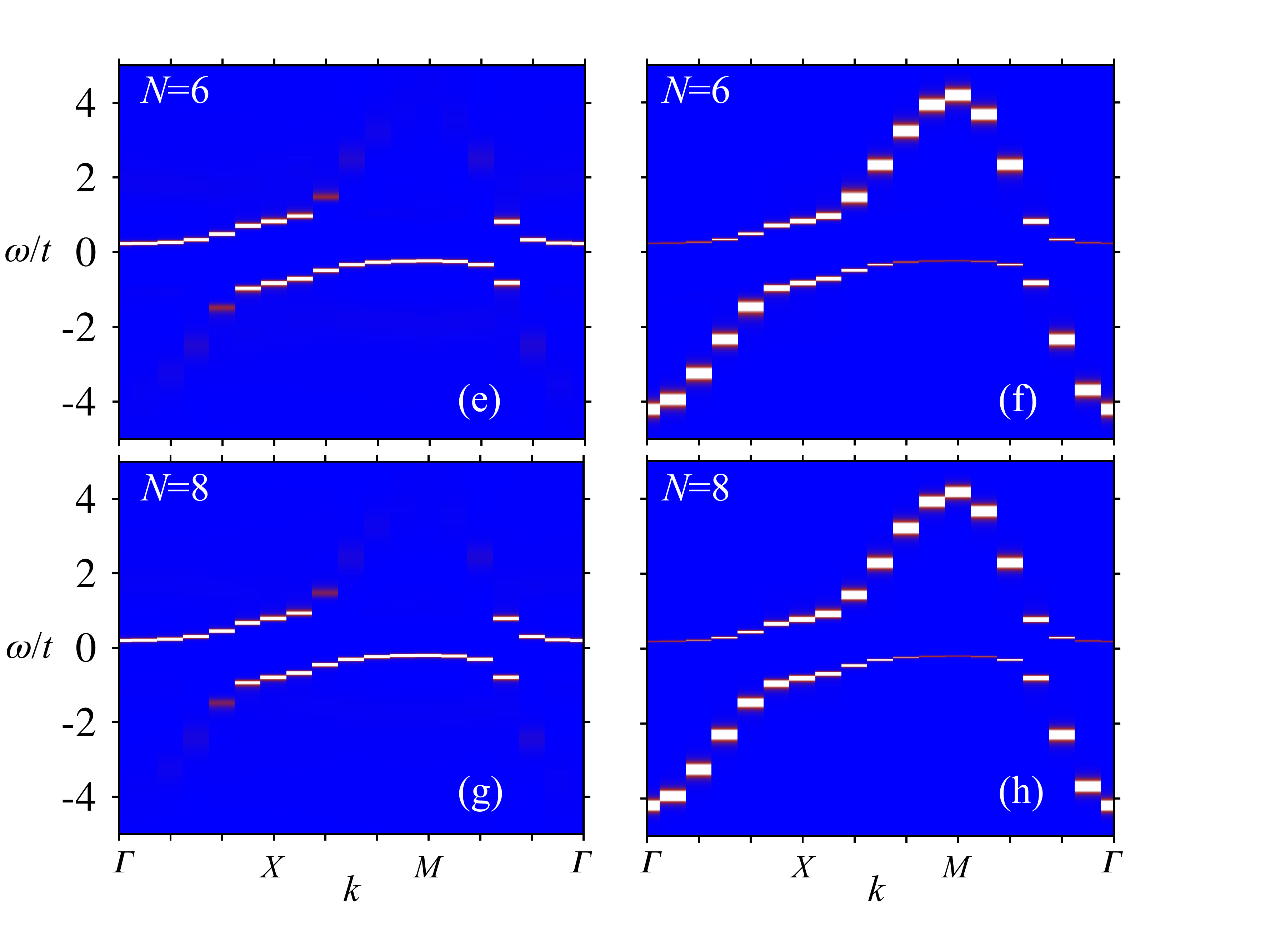}
\caption{Composite fermion $A_\psi(\k,\omega)$ (left) and conduction electron  $A_c(\k,\omega)$ (right)  spectral functions
at $T=0$ of the $12\times 12$  SU($N$) KLM in the Kondo-screened phase at $J_k/t=2$:
(a,b) $N=2$; (c,d) $N=4$; (e,f) $N=6$, and (g,h) $N=8$.}
\label{Jk2_SKLM_T0} 
\end{figure}

\begin{figure}[t]
\centering
\includegraphics[width=0.49\textwidth]{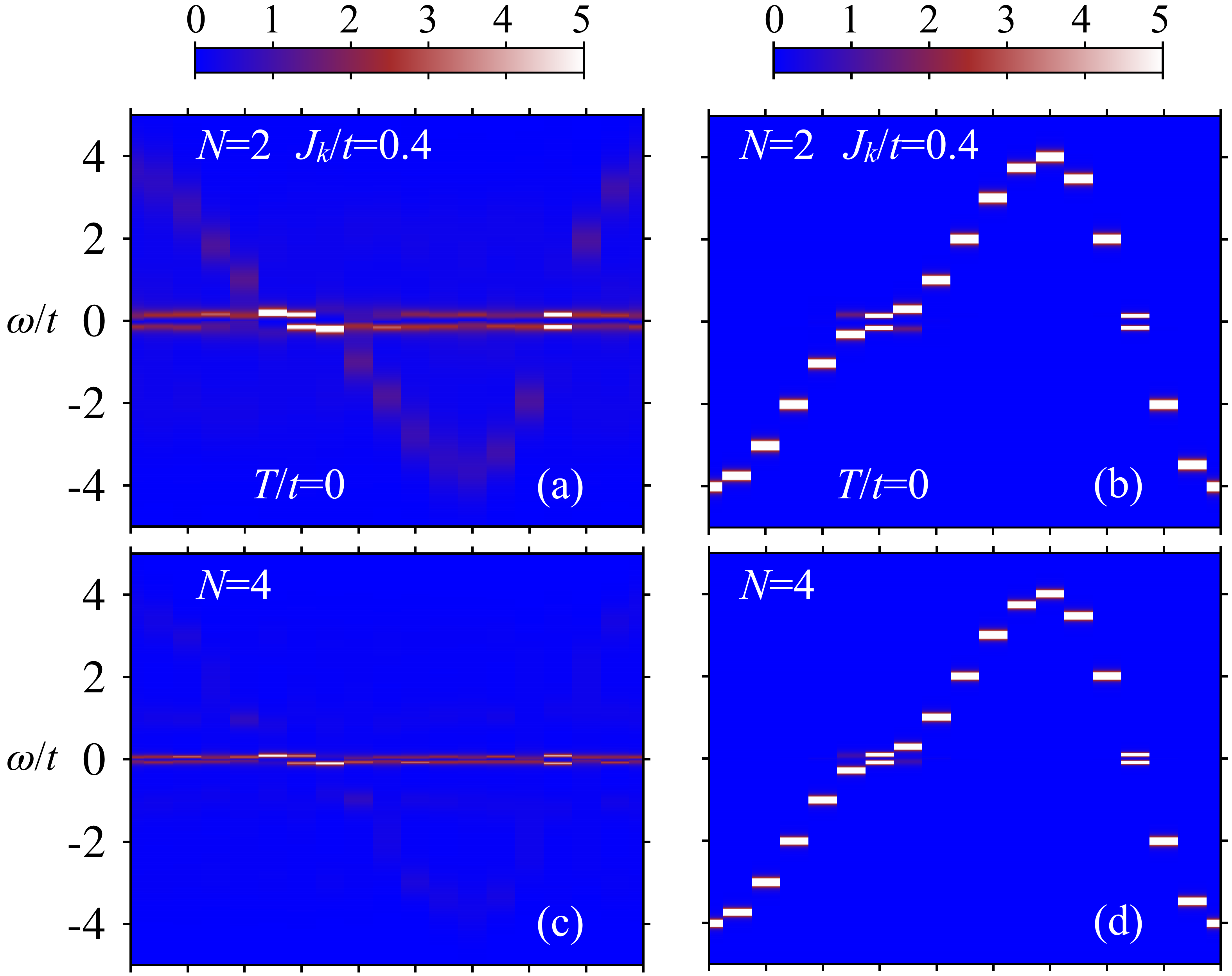}\\
\includegraphics[width=0.49\textwidth]{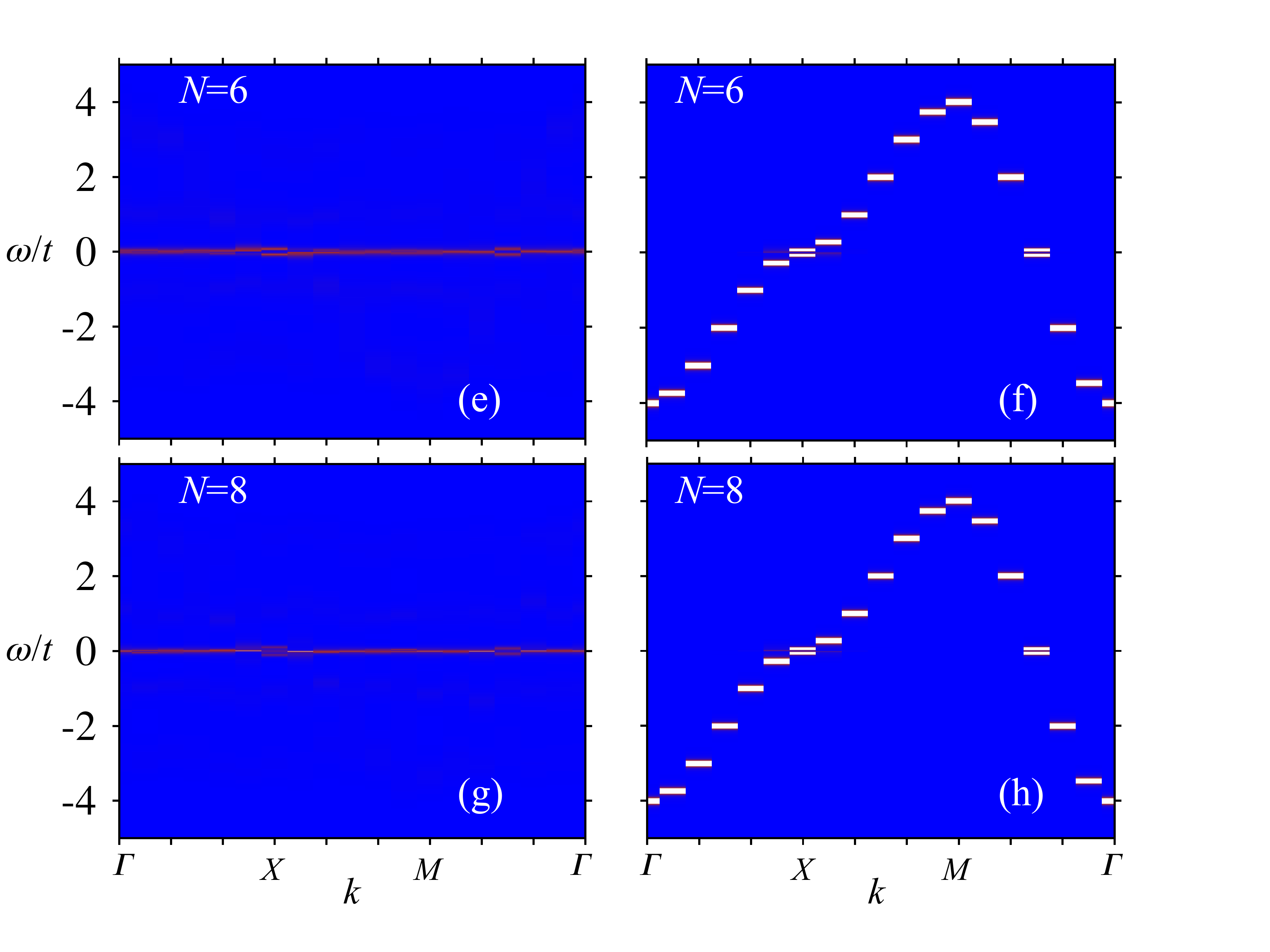}
\caption{Same as in Fig.~\ref{Jk2_SKLM_T0} but in the magnetically ordered phase at $J_k/t=0.4$.}
\label{Jk04_SKLM_T0} 
\end{figure}

We proceed now to discuss spectral properties of the composite fermion in the SU($N$) generalization of the KLM.  
Figures~\ref{Jk2_SKLM_T0}(a) and \ref{Jk2_SKLM_T0}(b) plot the zero temperature  $A_\psi(\k,\omega)$ and $A_c(\k,\omega)$ 
spectra for $N=2$ obtained in the Kondo insulating phase at $J_k/t=2$. Comparing with the corresponding finite-$T$ data, 
one easily recognizes the main spectral features whose momentum and frequency dependence as well as the intensity 
match well those found at $T=t/36$, see Figs.~\ref{Spectralcf_sq_L12_Beta36}(a) and \ref{Spectralcf_sq_L12_Beta36}(b). 
Furthermore, we note that increasing $N$  has a double effect, see Figs.~\ref{Jk2_SKLM_T0}(c)-\ref{Jk2_SKLM_T0}(h):  
(i) the quasiparticle gap is reduced, and (ii) overall, both the $A_\psi(\k,\omega)$ and $A_c(\k,\omega)$ spectra become
more coherent. This is a natural result given that larger $N$: (i) enhances the domain of stability of a magnetically 
disordered Kondo phase~\cite{Raczkowski20} bringing its description in line with a strong coupling picture, and 
(ii) reduces the effect of antiferromagnetic spin fluctuations on the self-energy such that it approaches the $\ve{k}$ 
independent large-$N$ limit.

We have equally plotted  both $A_\psi(\k,\omega)$ and $A_c(\k,\omega)$  at smaller values of $J_k/t=0.4$  as a function 
of $N$. This coupling strength is located in the phase diagram  deeply in the antiferromagnetically ordered phase.  
As evident in Fig.~\ref{Jk04_SKLM_T0}(a), the composite fermion spectrum $A_\psi(\k,\omega)$ at $N=2$ shares  
aspects of the large-$N$ and large-$S$ limits.  On the one hand, one observes  a low energy flat heavy fermion band 
with a renormalized weight accompanied by the shadow bands around the $\Gamma$ and $M$ momenta. These shadows display 
roughly the same intensity as the original heavy quasiparticles, thus reflecting robust magnetic order. 
On the other hand, a nearly fully polarized staggered magnetic moment results in a pronounced image 
of the $c$-band consistent with the large-$S$ limit. This image looses its intensity at larger $N$, see 
Figs.~\ref{Jk04_SKLM_T0}(c), \ref{Jk04_SKLM_T0}(e), and \ref{Jk04_SKLM_T0}(g). Seemingly, that could be 
a consequence of a reduced,  as a function of growing values of $N$, distance to the magnetic order-disorder 
transition point that scales as $J_c(N) \propto \frac{1}{\ln(N)}$~\cite{Raczkowski20}.
However, given that the $f$-local moment was found in Ref.~\cite{Raczkowski20}  to be next to saturated at $J_k/t=0.4$ 
for each considered $N$,  we find it more appropriate to assume that the loss in intensity originates from the scattering 
between a growing number  of Goldstone modes associated with the SU($N$) spin symmetry breaking in the N\'eel phase. 
This produces relatively broad spin wave excitations seen in the dynamical  spin structure  factor~\cite{Raczkowski20}, 
clearly beyond the lowest order approximation  in $S$ that leads us to a simplified form of the composite fermion Green's 
function in Eq.~(\ref{ferro_phaseQ}).  Strictly  speaking, large-$S$ theory  does not apply  here and one  should use  flavor wave  
theory as  done in Ref.~\cite{Kim_F17}.

In stark contrast to $A_\psi(\k,\omega)$, the corresponding conduction electron spectra $A_c(\k,\omega)$, see right panels of 
Fig.~\ref{Jk04_SKLM_T0}, display merely a direct hybridization gap without any discernible signature of the heavy fermion band. 
As we argue below, due to its extremely low intensity, the latter can only be resolved on a logarithmic scale.

\begin{figure}[t]
\centering
\includegraphics[width=0.48\textwidth]{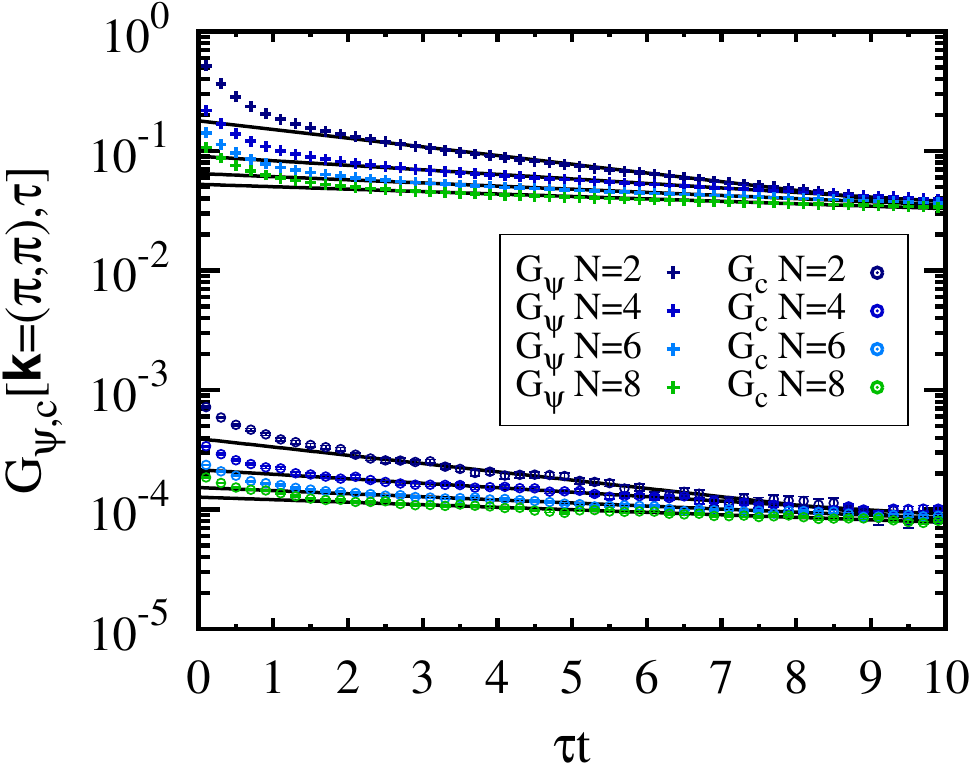}
\caption{Zero temperature composite fermion $G_{\psi}(\ve{k},\tau)$ and conduction electron $G_{c}(\ve{k},\tau)$ Green's functions  
at $\ve{k}=(\pi,\pi)$ in the SU($N$) KLM with $J_k/t=0.4$.
Both $G_{\psi}(\ve{k},\tau)$ and $G_{c}(\ve{k},\tau)$ follow at large values of $\tau t$ an exponential law 
$Ze^{-\Delta_{qp}\tau}$ (solid lines), thus indicating the existence of the heavy fermion band in their respective spectral 
functions. The latter have identical supports set by $\Delta_{qp}$  but differ, by nearly three orders of magnitude,  
in the quasiparticle weights $Z^{\psi}_{\ve{k}}$ and $Z^{c}_{\ve{k}}$.
}
\label{Green_SUN}
\end{figure}

To quantify the difference in quasiparticle weights of the heavy fermion band, we plot in Fig.~\ref{Green_SUN} 
raw data of the composite fermion $G_{\psi}(\ve{k},\tau)$ and conduction electron $G_{c}(\ve{k},\tau)$ Green's functions
at $\ve{k}=(\pi,\pi)$ obtained from QMC simulations of the SU($N$) KLM with $J_k/t=0.4$.
The quasiparticle residues $Z^{\psi}_{\ve{k}}$ and $Z^{c}_{\ve{k}}$ of the doped hole at momentum $\ve{k}$ 
can be estimated directly from the long-time behavior of the imaginary time Green's functions: 
\begin{align} 
\label{Psi_T0}
        G_{\psi}(\ve{k},\tau) &   
\stackrel{\tau \to \infty}{\to}  Z^{\psi}_{\ve{k}}e^{-\Delta_{qp}(\ve{k}) \tau }, \\
        G_{c}(\ve{k},\tau)    &  \stackrel{\tau \to \infty}{\to}  Z^{c}_{\ve{k}}e^{-\Delta_{qp}(\ve{k}) \tau }. 
\end{align}
As is apparent, both $G_{\psi}(\ve{k},\tau)$ and $G_{c}(\ve{k},\tau)$ show the same asymptotic behavior in the long-time 
limit, irrespective of $N$. It implies the continued existence of a pole in their respective spectral functions at the 
frequency $\omega=-\Delta_{qp}$, thus confirming the presence of the heavy fermion band. In contrast, the corresponding 
quasiparticle weights $Z^{\psi}_{\ve{k}}$ and $Z^{c}_{\ve{k}}$  are predicted  to differ by nearly three orders of 
magnitude, explaining the difficulty to resolve the heavy fermion band in $A_c(\k,\omega)$, see Fig.~\ref{Jk04_SKLM_T0}.

\begin{figure}[t]
\centering
\includegraphics[width=0.48\textwidth] {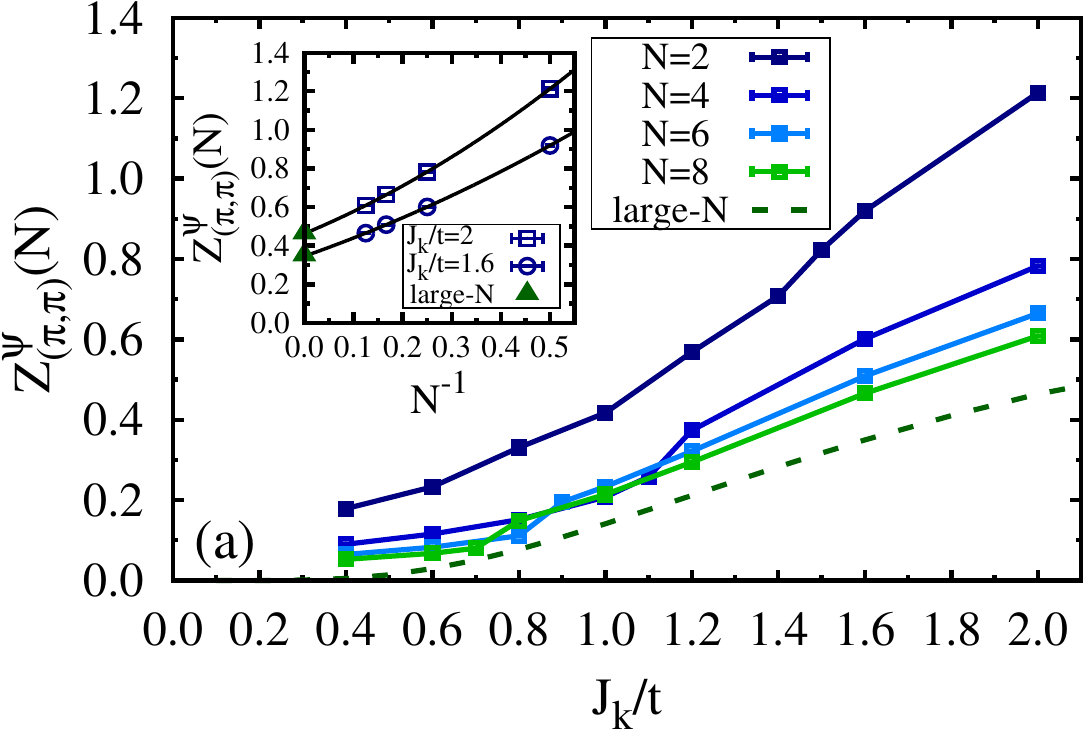}\\
\includegraphics[width=0.48\textwidth] {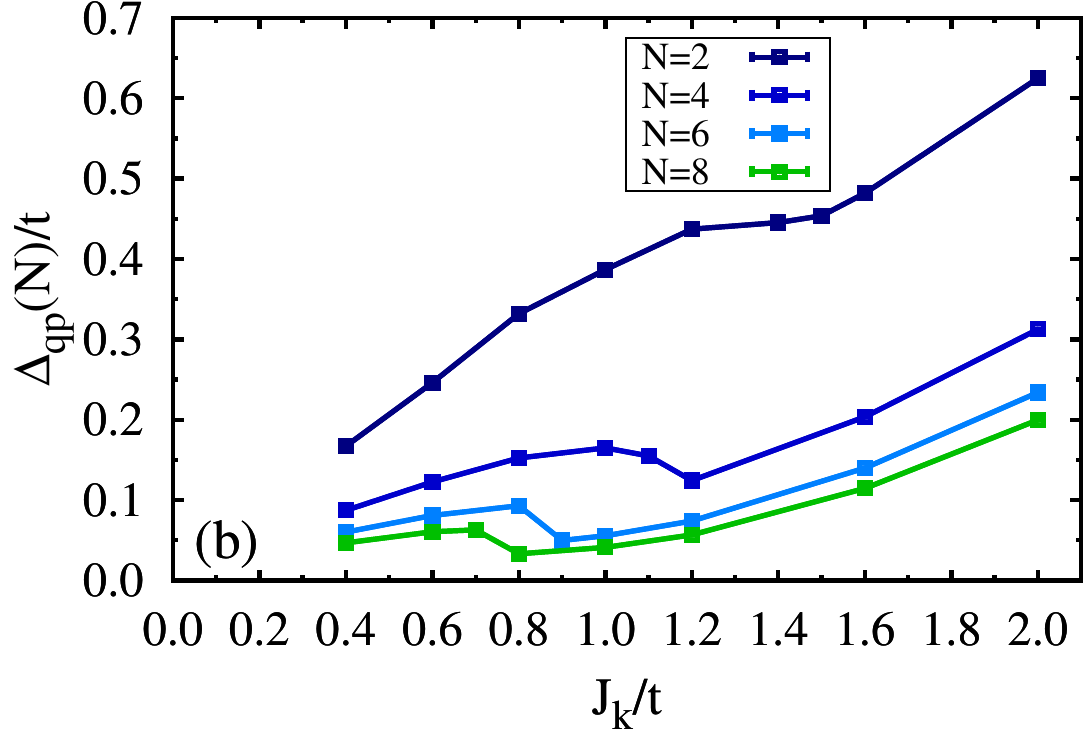}
\caption{(a) Quasiparticle residue $Z^{\psi}_{(\pi,\pi)}$ and (b) single particle gap $\Delta_{qp}$ of the composite fermion spectral 
function obtained from $T=0$ QMC simulations of the SU($N$) KLM. 
For comparison we also show $Z^{\psi}_{(\pi,\pi)}$ predicted by the large-$N$ approximation (dashed line).  
Inset shows a second-order polynomial fit to the QMC data in order to extract $Z^{\psi}_{(\pi,\pi)}$  
in the $N\to\infty$ limit; 
the extrapolated values  $Z^{\psi}_{(\pi,\pi)}(N\to\infty)= 0.463(4)$  at $J_k/t=2$ and
 $Z^{\psi}_{(\pi,\pi)}(N\to\infty)= 0.345(1)$ at $J_k/t=1.6$ 
match well those obtained using the large-$N$ approximation (triangles).  
}
\label{SUN_Zpsi}
\end{figure}

Next, we look at the evolution of the quasiparticle residue $Z^{\psi}_{(\pi,\pi)}$ as a function of $J_k/t$. 
We summarize it in Fig.~\ref{SUN_Zpsi}(a).
It conveys the main outcome of our study, i.e., numerical results for $Z^{\psi}_{(\pi,\pi)}$ in the $N=2$ case and in 
the large-$N$ approach exhibit an entirely different behavior in the small $J_k$ limit.  This is a counterintuitive result, 
since, as shown in  Eq.~(\ref{Veqpsi}),  the composite fermion operator directly provides the measure of hybridization, 
${\hat{\psi}}^{\dagger}_{\i,\sigma}\propto \frac{2}{N} {\hat{f}}^{\dagger}_{\i,\sigma} V$, where $V$ is the hybridization order 
parameter. Hence, one could expect $A_\psi(\k,\omega)\propto V^2 |u_{\k}|^2$, where $|u_\k|^2$ is the  coherence factor 
at the mean-field level, see Eq.~(\ref{coh_uk}). However, the spectral weight does not follow the exponentially small Kondo scale, 
displaying a linear dependency in $J_k/t$ instead, in analogy to the linear behavior of the  quasiparticle gap, see Fig.~\ref{SUN_Zpsi}(b). 
In the case of the quasiparticle gap, the linear dependency  was found to be a direct consequence of particle-hole symmetry 
and the associated Fermi surface nesting-driven magnetism~\cite{PhysRevB.82.245105,PhysRevB.96.155119,PhysRevB.98.245125}. 
The question then arises whether the same holds for the quasiparticle residue $Z^{\psi}_{(\pi,\pi)}$. 
Given  that the influence of the magnetism on $A_\psi(\k,\omega)$  boils mainly down to  
the backfolding of the heavy fermion band in the paramagnetic phase, we believe that the observed  enhancement of the spectral  weight  
is unrelated to the half-filled conduction band and stems instead from the specific form of the composite fermion operator.     

Another notable result is the abrupt reduction of $Z^{\psi}_{(\pi,\pi)}$  across the magnetic order-disorder transition point 
$J_c$ in the highly symmetric case with $N>2$.  
The ability of $Z^{\psi}_{(\pi,\pi)}$  to reflect the onset of long range magnetic order stems from the sum rule  for the   
composite fermion   spectral  function Eq.~(\ref{EQ.sum_rule}) which consists of the Kondo term 
$\langle  \hat{\ve{S}}_{\ve{i}}   \cdot \hat{\ve{c}}^{\dagger}_{\ve{i}}   \ve{\sigma} \hat{\ve{c}}_{\ve{i}}    \rangle$.   
This quantity was shown in Ref.~\cite{Raczkowski20} to display a discontinuous behavior at $J_c$ indicating a first-order 
nature of the transition for $N>2$. The latter is equally seen in the nonmonotonic behavior of the single particle gap, 
see Fig.~\ref{SUN_Zpsi}(b). 

Finally, by extrapolating finite-$N$ QMC data at $J_k/t=2$ and at $J_k/t=1.6$ to the $N\to\infty$ limit, 
we were able to recover the large-$N$ value of $Z^{\psi}_{(\pi,\pi)}$, see the inset in Fig.~\ref{SUN_Zpsi}(a).  
This confirms that the large-$N$ theory is the correct saddle point of the SU(2) KLM in the Kondo regime.

\subsection{Ferromagnetic  Kondo lattice} 

 \begin{figure}[t]
\centering
\includegraphics[width=0.49\textwidth]{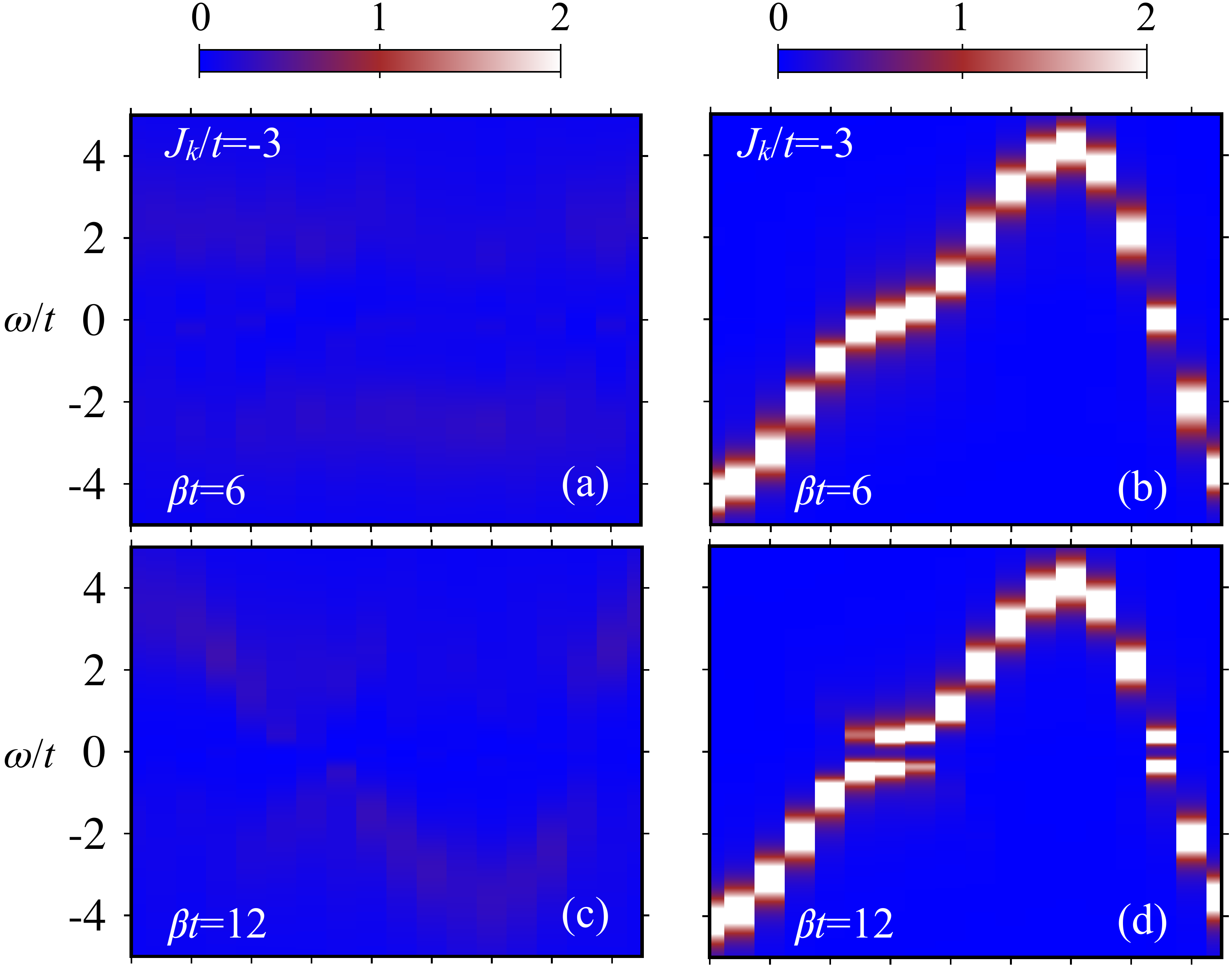}\\
\includegraphics[width=0.49\textwidth]{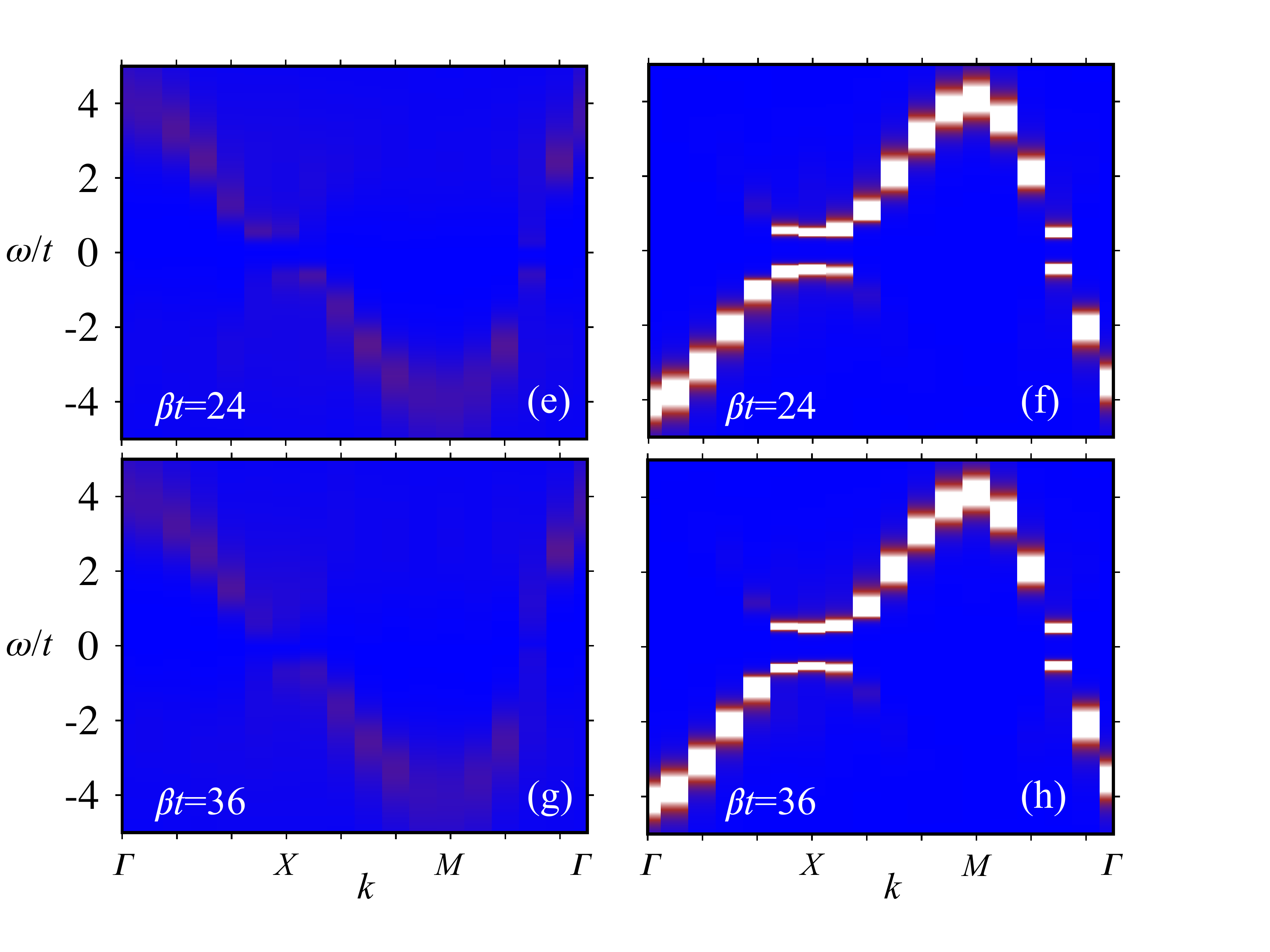}%
\caption{Temperature dependence of  $A_\psi(\k,\omega)$ (left) and $A_c(\k,\omega)$ (right) in the $12\times 12$ KLM with ferromagnetic coupling $J_k/t=-3$: 
(a,b) $\beta t=6$;  (c,d) $\beta t=12$;  (e,f) $\beta t=24$, and (g,h) $\beta t=36$. }
\label{Ferro_spec_sq_L12_Beta} 
\end{figure}

\begin{figure}[t]
\centering
\includegraphics[width=0.45\textwidth]{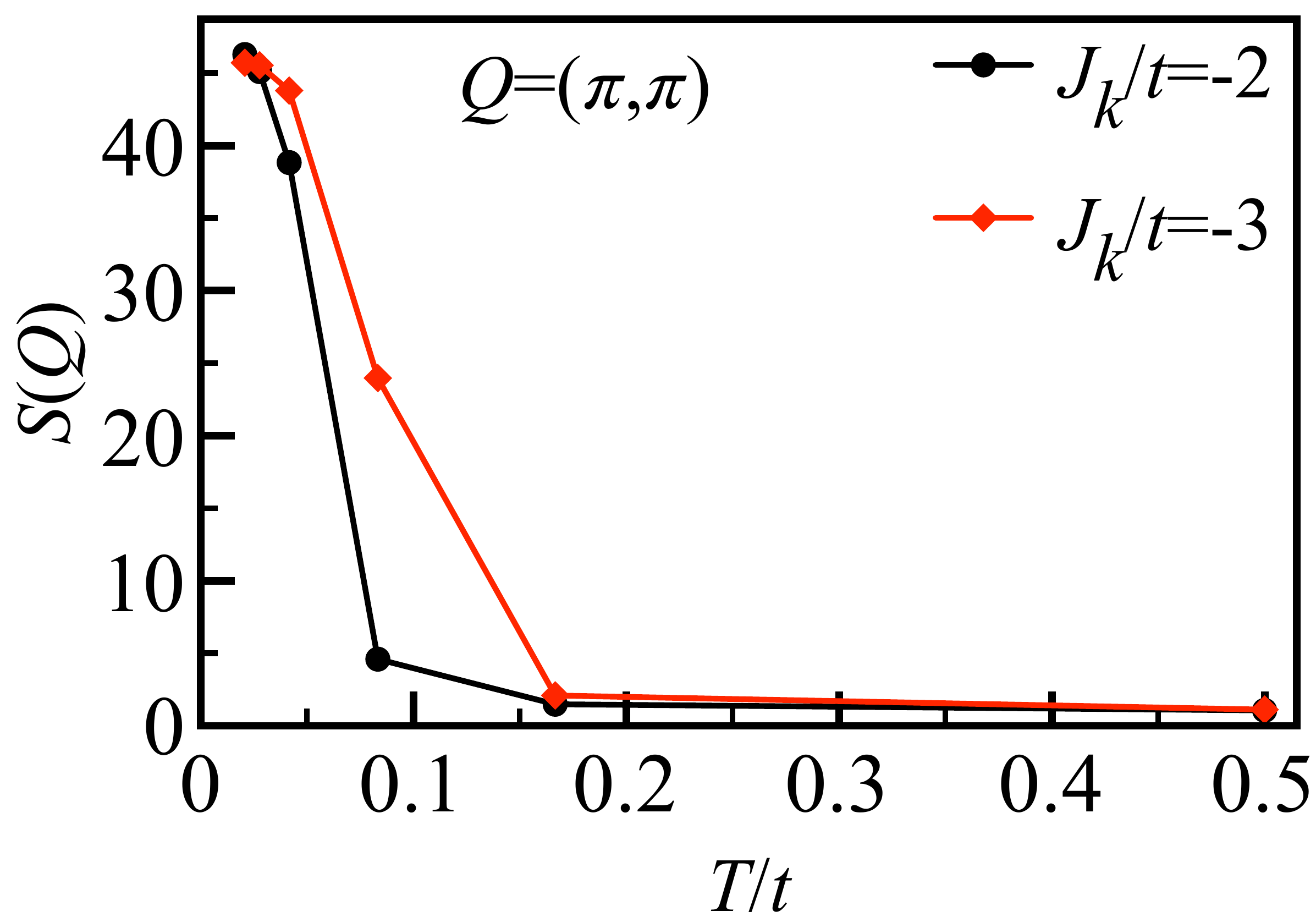} 
\caption{Static spin structure factor $S(\Q)$ at $\Q=(\pi,\pi)$ for the localized spins as a function of temperature $T/t$ in  the  ferromagnetic KLM. }  
\label{SQf_vs_T_SKLM_FM} 
\end{figure}

The  ferromagnetic KLM offers the possibility to retain the  RKKY  interaction  and    switch off the Kondo effect.    
Fig.~\ref{Ferro_spec_sq_L12_Beta}    plots the  temperature dependence of the  composite  fermion  and  conduction electron 
spectral functions as a function of temperature  at $J_k/t = -3$.   The relevant  energy scale  required to interpret the plots is  the magnetic
scale  below which antiferromagnetic fluctuations set in.  From Fig.~\ref{SQf_vs_T_SKLM_FM}  we   can  estimate   $T_{RKKY} \simeq  0.125 t $.  
At   $  T  \ll T_{RKKY}   $, see Figs.~\ref{Ferro_spec_sq_L12_Beta}(e)-\ref{Ferro_spec_sq_L12_Beta}(h),   one finds that the  data are well  reproduced by  
 the  large-$S$  results.   In particular   the  single particle  Green's functions match well  the forms: 
\begin{eqnarray}
A_c(\k,\omega)   =  &  &  \frac{1}{2} \left(   1 +   \frac{\epsilon(\ve{k}) }{E(\ve{k})} \right)  \delta( \omega -  E(\ve{k}) )  +   \nonumber  \\
   & &  \frac{1}{2} \left(   1 -   \frac{\epsilon(\ve{k}) }{E(\ve{k})} \right)  \delta( \omega + E(\ve{k}) ). 
\end{eqnarray}
Here $\epsilon(\ve{k})=   -2 t\left(  \cos{k_x} +  \cos{k_y} \right) $   is  the  noninteracting dispersion relation and 
$ E(\ve{k})  = \sqrt{\epsilon(\ve{k})^2  + \Delta^{2}} $,  with $\Delta$  the gap.     
Although the   dispersion relation is  independent of the  shift  of the ordering wave   vector  $\ve{Q}   =(\pi,\pi)$,  the   coherence  factors  are not.    
Hence, the  dominant weight  in  $A_c(\k,\omega)$  [$A_\psi(\k,\omega)$]   follows  the   noninteracting dispersion relation  
$\epsilon(\ve{k})$   [$ \epsilon(\ve{k} + \ve{Q} ) $].     This is consistent with the large-$S$  relation of 
Eq.~(\ref{ferro_phaseQ})   corresponding  to  $ A_\psi(\k,\omega) \simeq  S^2 A_c(\k+\ve{Q},\omega) $.  

At high temperatures,   $T >  T_{RKKY} $, the magnetically  induced  gap should vanish and,  as   argued in  Eq.~(\ref{Apsieq_kinpndt}),  
the   composite fermion spectral  function is expected to show  no $\ve{k}$-dependence. 
Figures~\ref{Ferro_spec_sq_L12_Beta}(a) and \ref{Ferro_spec_sq_L12_Beta}(b) confirm the above expectations.

The above   demonstrates  that the  dominant effects of the RKKY interaction can  be understood in terms of  a large-$S$ or  mean-field approximation.

\section{Summary and Conclusions}\label{Conclusions}

The composite fermion operator  we  have considered in this article is  defined in Eq.~(\ref{composite_fermion})   and is   at  best understood  in terms of a canonical  Schrieffer-Wolff   transformation   of the  electron creation operator in  a  localized Wannier state of the PAM.    We have  studied  
numerically the  spectral  function of the  composite fermion for the   SU($N$) antiferromagnetic  and  SU(2)  ferromagnetic KLM on a square  lattice  and  
provided a number of insights   based on analytical considerations in the large-$N$,  large-$S$ and   strong  coupling limits.

The key   result of the paper is   numerical.   We observe that   for the  antiferromagnetic SU(2)  KLM on the  square lattice   the  spectral function of 
the composite fermion,  $A_{\psi}(\ve{k}, \omega) $,  reveals  the  heavy  fermion  band and  that  the spectral weight  tracks   $J_k/t$ in the weak coupling 
limit.     This  should be contrasted with the conduction electron spectral function,   $A_{c}(\ve{k}, \omega) $, that also  captures the   heavy  fermion   
band but    with spectral weight   given by the Kondo  scale  $e^{-W/J_k}$.  Hence  the composite fermion provides a  remarkable  enhancement of  spectral  
intensity  in the weak coupling limit  and   greatly facilitates investigations  of the  heavy fermion band in the realm of the  SU(2) KLM.  

 For model Hamiltonians  that   support   quasiparticle excitations,    we   generically  expect the  dispersion relation to  be  revealed by the  spectral  
 function of  \textit{ any } operator  with appropriate quantum numbers.   Thereby, the effective mass  of charge carriers   --  corresponding to  the  
 inverse curvature of the  dispersion relation --   is independent  of the choice of   the fermion operator.     In terms  of the  self-energy,  the effective 
 mass is given by the inverse  quasiparticle  residue --  that   stems from its    frequency dependency   --    times  a  term that reflects  
 its  momentum dependence~\cite{Negele}.    Hence  $m^* =  m^*_{\ve{k}} /  Z_{\ve{k}} $.   
 Let us apply  the  above to the  SU(2) KLM.   Here $m^{*} \propto e^{W/J_k}  $ in the   weak coupling limit~\cite{Assaad04a}.  Since   
$Z^{\psi}_{\ve{k}=(\pi,\pi)} \propto J_k$ we conclude that it is the  $\ve{k}$-dependence of the  
 composite fermion self-energy that   captures the  heavy fermion effective mass.    More   generally,  optimization of the   quasiparticle  residue  
 necessitates real space   fluctuations.  
 Both in the large-$N$  limit~\cite{Burdin00}   and at $N=2$~\cite{Assaad04a},   the effective  mass tracks  the inverse  Kondo temperature.  
 However, the  quasiparticle  residue of the composite fermion varies from   $Z_{\ve{k}}    \propto e^{-W/J_k} $   in the large-$N$ limit  to $Z_{\ve{k}}    \propto J_k $   at $N=2$   thereby reflecting the buildup of  spatial   fluctuations as a function of    decreasing $N$.   
 Clearly  as a function of  $N$,   and  at sufficiently small  values of $J_k$,    we will encounter a  magnetic  phase transition~\cite{Raczkowski20}  such that 
 the question  arises   if the  observed   enhancement of the spectral  weight  is a consequence of  magnetic fluctuations.   We  believe that this is not the 
 case    since the spectral function of the composite  fermion is  very well  understood  in terms  of backfolding of the heavy fermion band in the 
 paramagnetic phase  and  concomitant opening of  a  single  particle gap set by $J_k$    reflecting the particle-hole   symmetry.    
Accordingly, the composite fermion opens up the possibility  to track the fate of the fragile heavy fermion quasiparticle in the KLM in any situation  
where  competing instabilities lead to a significant suppression of the lattice Kondo effect and thus to extremely 
low coherence temperatures~\cite{PhysRevB.81.054444,Bodensiek2011,PhysRevB.98.195111}.

The aforementioned  growth of  spectral  weight between the large-$N$ and SU(2)  limits  can also be seen  when building a  KLM  by assembling   magnetic  adatoms 
on a  metallic  surface~\cite{Raczkowski18}.   
In the  single impurity limit the   composite  fermion   local spectral function  reveals the Kondo  resonance  with spectral  weight tracking  the Kondo  scale.   
As  magnetic adatoms are assembled  around this  initial  impurity so as to locally form  a half-filled  Kondo lattice,   the Kondo  resonance   develops a  gap and 
acquires substantial  spectral weight.  This is explicitly seen in Fig. 1(c) of  Ref.~\cite{Raczkowski18}.

In the  strong  coupling limit,   the Kondo effect  corresponds to the formation  of a singlet  between the  conduction electron and  spin degree 
of  freedom in a unit cell.  In this limit the half-filled  ground state, $|\Psi_0^{n} \rangle $, of the  SU(2)  KLM   is  a direct product   of  such  
singlets. As  apparent from Eqs.~(\ref{c_bond.eq})   and (\ref{psi_bond.eq}),   
$   \hat{c}_{\ve{i},\uparrow} | \Psi_0^n \rangle =  - 2  \hat{\psi}_{\ve{i},\uparrow} | \Psi_0^n \rangle  $.   
Hence in this limit,   we observe,  up to a normalization factor,  no  difference  between the $c$-  and $\psi$-  spectral functions.    The  strong  coupling 
limit  is  characterized by fast   magnetic fluctuations  on the time scale of the motion of a doped electron. Thereby, the  picture of  a doped hole moving  
in a Kondo singlet  background is  appropriate.  In the weak coupling limit,   magnetic fluctuations  are  slow in comparison to the  hole motion,  such that 
locally  the  doped hole will   perceive  a static  magnetic background.    In this case  our results show that  
$ \hat{c}_{\ve{k},\uparrow} | \Psi_0^n \rangle  $ and $ \hat{\psi}_{\ve{k},\uparrow} | \Psi_0^n \rangle $ differ substantially.  In  particular,  
$ \hat{\psi}_{\ve{k},\uparrow} $    removes a conduction electron   and  simultaneously  \textit{ adjusts } the  spin  background  whereas  
$ \hat{c}_{\ve{k},\uparrow} $    merely    destroys a  conduction electron.  Our  result suggests that  
$ Z^{\psi}_{\ve{k} = (\pi,\pi)} / Z^{c}_{\ve{k} = (\pi,\pi)} $   diverges in the  weak coupling such  that   $\hat{\psi}_{\ve{k},\uparrow} $  captures the 
\textit{    correct }   form of the  heavy fermion quasiparticle.  
In particular  it  corresponds to a bound state of  conduction electron and spin degrees of freedom.    The  energy scale  of the bound state is  revealed by  
$A_{\psi}(\ve{k},\omega) $, see Fig.~\ref{Spectralcf_sq_L12_Beta36},   as  the energy scale  above   which the  quasiparticle pole \textit{ dissolves }  into a  continuum  captured by the convolution of the  spin  susceptibility and   spectral  function of the conduction electron.   To  a first approximation,    
our  results  suggest that this energy  scale  tracks  $J_k$.

$A_{\psi}(\ve{k}, \omega) $   is a quantity of  choice to study  Kondo breakdown transitions  since  at this  transition we  expect  
the  destruction of the aforementioned   bound state.    This statement  is   supported by our large-$N$  results  where the composite fermion
operator    reveals the   hybridization  matrix element   that vanishes at a  Kondo breakdown transition.  
It   furthermore  follows  from the very  definition of the  composite fermion operator in terms of a Schrieffer-Wolff  transformation of the electron creation operator in a  localized orbital in  the PAM.  In particular,  in the realm of the PAM and in the Kondo breakdown phase,  the localized electrons drop   out  from the low  energy  physics. 
 For Kondo breakdown critical points in metallic  environments  we hence expect $A_{\psi}(\ve{k}, \omega) $    to develop a  gap  thereby  revealing the   orbital selective  Mott  nature of this transition \cite{Vojta10}.   This has already  been partially  observed in Ref.~\cite{Danu20}.    However,  even  in an  insulating state, Kondo breakdown  should correspond to the destruction of the   low lying quasiparticle pole in $A_{\psi}(\ve{k}, \omega) $. The  fact  that we do  not observe this for the square lattice,  points to the  fact that Kondo  screening  and magnetism coexist down  to  our lowest  considered value of $J_k/t=0.4$.

The  above can be checked by considering the  ferromagnetic Kondo lattice  where the Kondo effect  is absent.  Here, we  have seen that the spectral function 
shows no low lying  bound states    and that it can be very well  understood    within a large-$S$  expansion: at  leading order in $S$   the composite  
fermion spectral  function  reduces to  the conduction one shifted by  the  momentum $\ve{Q} $  of the magnetic ordering.

The   aforementioned   derivation of the composite fermion operator   from the PAM in the local  moment  regime, allows us to  link our results 
to  experiments.
The local composite  fermion   spectral function  can be  measured  with STM  experiments of  adatoms on metallic  surfaces  provided  that the  current  
between the  STM tip and metallic surface   flows   through the  correlated orbital of the adatom. This has been  achieved   by  capping a  metallic  surface   
with an  insulating layer  on which  adatoms reside~\cite{Toskovic2016}.    Photoemission  studies of the 4$f$  levels  of  CePt$_5$  surface   alloys  have been 
presented in Ref.~\cite{Klein08}.  Since this compound is in the local moment  regime, one can conjecture that the appropriate  spectral function   required 
to  capture the heavy fermion quasiparticle is   $A_{\psi}(\ve{k}, \omega) $  in the framework of a  Kondo lattice modeling.    Our  results suggest that  the 
$\ve{k}$-dependence of the  self-energy plays  an important  role for the  understanding of $A_{\psi}(\ve{k}, \omega) $.   This  may  explain  why   dynamical 
mean-field calculations in combination with a  non-crossing approximation  of the  spectral function in the realm of the  PAM   equally  presented in 
Ref.~\cite{Klein08},  seem to  underestimate the  spectral    weight of the heavy fermion bands   in comparison to experiments.

\begin{acknowledgments}
The authors gratefully acknowledge the Gauss Centre for Supercomputing
e.V. (www.gauss-centre.eu) for funding this project by providing computing
time on the GCS Supercomputer SUPERMUC-NG at Leibniz Supercomputing Centre
(www.lrz.de) as well as through the John von Neumann Institute for Computing (NIC) on the GCS Supercomputer JUWELS  \cite{JUWELS} at the 
J\"ulich Supercomputing Centre (JSC).  BD and ZL  thank the W\"urzburg-Dresden Cluster of Excellence on
Complexity and Topology in Quantum Matter ct.qmat (EXC 2147, project-id
390858490) for  financial support.   FFA thanks  support from the DFG funded SFB 1170
on Topological and Correlated Electronics at Surfaces and Interfaces.
\end{acknowledgments}

\bibliography{fassaad,ref,marcin}
 \end{document}